\documentclass[12pt,letterpaper]{article}
\pdfoutput=1

\usepackage{graphicx,array}
\usepackage{color}
\usepackage{latexsym}
\usepackage{amsthm}
\usepackage{amsmath}
\usepackage{amssymb}
\usepackage[numbers,sort&compress]{natbib}
\usepackage{bm}
\usepackage{slashed}
\usepackage{mathrsfs}
\usepackage{hyperref} %Automatically links \label and \ref commands; Always load last
\hypersetup{
    colorlinks=true,       % false: boxed links; true: colored links
    linkcolor=red,          % color of internal links
    citecolor=blue,        % color of links to bibliography
    filecolor=magenta,      % color of file links
    urlcolor=blue           % color of external links
}
\usepackage[all]{hypcap} %Link navagates to top of figure instead of caption (below fig)

\usepackage{myterms}
\usepackage{latexsym}

\newcommand{\dQN}{\mathrm{dQN}}
\newcommand{\dQM}{\mathrm{dQM}}
\newcommand{\B}{\textsf{B}}
\newcommand{\pvec}{{\bm p}}
\newcommand{\UV}{\text{\sc uv}}

\newcommand{\Bd}{{\mathsf{B}_d}}
\newcommand{\dB}{{B_d}}
\newcommand{\Tc}{T_{\gamma,c}}
\newcommand{\nc}{\newcommand}
\nc{\beq}{\begin{equation}}
\nc{\eeq}{\end{equation}}
\nc{\beqa}{\begin{eqnarray}}
\nc{\eeqa}{\end{eqnarray}}
\nc{\bit}{\begin{itemize}}
\nc{\eit}{\end{itemize}}

\numberwithin{equation}{section}

\title{\bf  Dark Quark Nuggets}

\author{\large Yang Bai$^{a,b}$, Andrew J. Long$^{c,d}$, and Sida Lu$^{a}$}
\date{\small \it 
$^a$Department of Physics, University of Wisconsin-Madison, Madison, WI 53706, USA \\
$^b$Theoretical Physics Department, Fermilab, Batavia, IL 60510, USA \\
$^c$Kavli Institute for Cosmological Physics, University of Chicago, Chicago, Illinois 60637, USA 
$^d$Leinweber Center for Theoretical Physics, University of Michigan, Ann Arbor, Michigan 48109, USA 
}

\begin{document}

\maketitle

\setlength{\parskip}{0.2ex}

\begin{abstract}
``Dark quark nuggets'', a lump of dark quark matter, can be produced in the early universe for a wide range of confining gauge theories and serve as a macroscopic dark matter candidate. The two necessary conditions, a nonzero dark baryon number asymmetry and a first-order phase transition, can be easily satisfied for many asymmetric dark matter models and QCD-like gauge theories with a few massless flavors. For confinement scales from 10 keV to 100 TeV, these dark quark nuggets with a huge dark baryon number have their masses vary from $10^{23}~\mathrm{g}$ to $10^{-7}~\mathrm{g}$ and their radii from $10^{8}~\mathrm{cm}$ to $10^{-15}~\mathrm{cm}$. Such macroscopic dark matter candidates can be searched for by a broad scope of experiments and even new detection strategies. Specifically, we have found that the gravitational microlensing experiments can probe heavier dark quark nuggets or smaller confinement scales around 10 keV; collision of dark quark nuggets can generate detectable and transient electromagnetic radiation signals; the stochastic gravitational wave signals from the first order phase transition can be probed by the pulsar timing array observations and other space-based interferometry experiments; the approximately massless dark mesons can behave as dark radiation to be tested by the next-generation CMB experiments; the free dark baryons, as a subcomponent of dark matter, can have direct detection signals for a sufficiently strong interaction strength with the visible sector.
\end{abstract}

\newpage

\begingroup
\hypersetup{linkcolor=black}

\hypersetup{linktocpage} % the table of content can have hyperlink
\tableofcontents 
\hypersetup{linkcolor=red} %after fix black color for table of content and change the link color

\endgroup

\newpage

%==================================
% Introduction
%==================================
\section{Introduction}\label{sec:introduction}

%=========
The theory of quantum chromodynamics (QCD) is an integral part of the Standard Model (SM) of elementary particles as it successfully explains hadron properties, nuclear structure and phenomena.  
While QCD predicts that most matter in the current universe is in the form of hadrons, the theory also admits an exotic phase of ``quark matter'' at high baryon-number density and low temperature~\cite{Alford:2007xm}.  
In his seminal work, Witten~\cite{Witten:1984rs} proposed that ``nuggets'' of quark matter could have formed in the early universe at the epoch of quark confinement, and that these nuggets could survive in the universe today as a dark matter candidate.  
One can understand Witten's quark nuggets as macroscopic nucleons (not nuclei) with a very large baryon number, $N_\B > 10^{30}$.  
Whereas Witten assumed that our QCD confining phase transition was a first order one, numerical lattice studies later revealed that the transition is predicted to be a continuous crossover instead~(see e.g., \rref{Fodor:2001pe}), and therefore quark nugget production is not viable in the SM.  

%=========
Nevertheless, the requirements for quark nugget production are generic, and although SM QCD does not have all the right ingredients, it is not hard to find new physics, beyond the Standard Model (BSM), that facilitates the formation of these objects.  
In particular, the formation of nuggets needs  {\it i}) a first-order phase transition to have (at least) two phases with different vacuum energies; {\it ii}) a conserved global charge for a small pocket of space to build up a large global charge;  {\it iii}) a cosmological excess of matter over antimatter, corresponding to a nonzero density of a conserved global charge.  
The SM QCD satisfies the last two conditions but not the first one.
Regarding the first condition, the literature on BSM physics is replete with confining gauge theories including the UV-completion of composite Higgs model~\cite{Kaplan:1983sm,Nussinov:1985xr}, supersymmetric models~\cite{Intriligator:1995au,Bergner:2014saa}, Twin Higgs models~\cite{Chacko:2005pe}, dark QCD~\cite{Bai:2010qg,Bai:2013xga,Boddy:2014yra,Hochberg:2014dra,Appelquist:2015yfa,Antipin:2015xia,Kribs:2018oad,Kribs:2018ilo} and Nnaturalness models~\cite{Arkani-Hamed:2016rle}.  
As we will discuss further in \sref{sec:dark_QCD}, the condition of a first order phase transition is easily satisfied as long as the number of light vector-like fermions obeys $N_f \geq 3$ for an $\SU{N}$ gauge theory.  
(In SM QCD the up and down quarks are light compared to the confinement scale, but the strange quark is marginal, and consequently the QCD phase transition is not first order.)  
For the second condition and similar to the $\U{1}$ baryon number in the SM, it is natural to have (approximately) good symmetry in the new strong-dynamics sector such as technibaryon, twin baryon, and dark baryon number symmetries.  
Finally, for the third condition it is natural to expect that a matter-antimatter asymmetry may be shared between the dark and visible sectors~\cite{Dick:1999je,Murayama:2002je,Shelton:2010ta,Buckley:2010ui,Haba:2010bm,Davoudiasl:2010am,Blennow:2010qp,Allahverdi:2010rh,Ibe:2018juk}. 

%=========
It is interesting to remark here that, based on the conditions above, the presence of dark quark nuggets may be unavoidable in some models of dark baryon dark matter~\cite{Nussinov:1985xr}. As we will discuss in Sec.~\ref{sec:dark_QCD}, for models with three or greater flavors of light dark quarks, the confining phase transition is expected to be a first order one, and dark quark nuggets can be formed. The dark baryon number could be mainly in the dark quark nugget states, similar to the QCD nuggets in Ref.~\cite{Witten:1984rs}. This observation motivates a reevaluation of earlier studies of dark baryon dark matter to assess whether those models also predict a relic abundance of dark quark nuggets.

%=========
In this work we consider a class of BSM confining gauge theories, collectively denoted as ``dark QCD,'' which are parametrized by the number of colors, the number of flavors of light vector-like fermions, and the confinement scale.  
We study the properties of ``dark quark matter'' and the conditions under which stable ``dark quark nuggets'' (dQN) can form through a cosmological phase transition in the early universe.  
Depending on the confinement scale, the typical nugget's mass and radius can reach as large as $M_\dQN \sim 10^{23} \gram$ and $R_\dQN \sim 10^8 \cm$. We argue that these nuggets can survive in the universe today where they provide a candidate for the dark matter, and we explore various observational prospects for their detection.  

%=========
Dark quark nuggets are examples of macroscopic dark matter; for a recent review see \rref{Jacobs:2014yca}. Given the null results of searching for weakly interacting massive particle with a mass of $O(100\,\mbox{GeV})$~\cite{Aprile:2018dbl}, it is natural to explore other well-motivated dark matter models with different masses.
 Since the last several years have seen renewed interests in these dark matter candidates, let us briefly note some of the recent developments and clarify their connection to our own work.  
To our knowledge the author of \rref{Witten:1984rs} was the first to propose that the dark matter could consist of macroscopic objects with nuclear densities, and he called these objects {\it quark nuggets} since they were made up of Standard Model quark matter. 
Subsequent work introduced a coupling to the QCD axion, which led to {\it axion quark nuggets}, where quark nuggets are formed through CP-violating domain walls with modified properties and enhanced stability~\cite{Zhitnitsky:2002qa,Lawson:2012zu,Atreya:2014sca}.
Other authors proposed that {\it six-flavor quark nuggets} could form if the electroweak phase transition were supercooled to the QCD scale~\cite{Bai:2018vik}.  
The nuggets that is made of techniquarks have also been studied in technicolor models~\cite{Frieman:1990nh}.
%\SL{The idea of quark nugget was also borrowed into technicolor models soon after its proposal~\cite{Frieman:1990nh}.}

%=========
The more recent interest in macro dark matter is motivated by the idea that dark matter's self-interactions can allow composite objects to form by aggregation.
Several authors have considered that the dark sector could undergo a period of {\it dark nucleosynthesis} to form composite objects with $O(1)$ constituents~\cite{Krnjaic:2014xza,Detmold:2014qqa,Hardy:2014mqa,McDermott:2017vyk}.  
The authors of \rrefs{Wise:2014ola,Wise:2014jva} studied a model of asymmetric dark matter in which $O(\gg 1)$ Dirac fermions become bounded together through a Yukawa interaction via a light scalar mediator and form a non-relativistic degenerate Fermi gas; they called these objects {\it dark matter nuggets}.  
In work by other authors, the properties and production mechanism of these {\it asymmetric dark matter nuggets} was clarified and refined~\cite{Gresham:2017zqi,Gresham:2017cvl}. The authors of \rref{Grabowska:2018lnd} considered composite objects, which they called {\it dark blobs}, that can be formed from either bosonic and fermionic constituent particles, and they study the associated detection strategies.

%=========
The remainder of this article is organized as follows.  
In this work we study a class of BSM confining gauge theories, collectively denoted as ``dark QCD,'' that are introduced in \sref{sec:dark_QCD}.  
We discuss the conditions under which the confining phase transition is a first order one, which is a necessary condition for the formation of dark quark nuggets.  
In \sref{sec:dark_quark_matter} we analyze the properties of dark quark matter and discuss how the Fermi degeneracy pressure provided by the (conserved) dark baryon number supports the dark quark nugget against collapse.  
\sref{sec:dark_nuggets} address the cosmological production of dark quark nuggets and contains estimates for their mass, size, and cosmological relic abundance.  
In \sref{sec:signatures} we discuss various observational signatures including gravitational wave radiation, dark radiation, colliding and merging signatures, and prospects for direct detection.  
We conclude in \sref{sec:conclusion}.  
In \aref{app:EFT}, we provide a calculation of the phase transition based on the effective sigma model for the dark chiral symmetry breaking.

%==================================
% Dark quantum chromodynamics
%==================================
\section{Dark quantum chromodynamics}\label{sec:dark_QCD}

%=========
In this section we introduce the model being considered in the remainder of the article.  
In particular we are interested in ``dark QCD'' with $N_d$ colors and $N_f$ flavors of (approximately massless) vector-like fermions. In our model, we will assume that there is no dark electroweak gauge group or dark neutrino. More or less, the dark QCD is anticipated to have a similar asymptotic-free dynamics as our SM QCD. In an ultra-violet energy range, the dark $\SU{N_d}$ QCD has a perturbative gauge coupling and with the particle content composed of $N_d^2 - 1$ dark gluons, $N_f$ dark quarks, and $N_f$ dark antiquarks. The gauge coupling becomes strong in an infrared scale $\Lambda_d$ and both confinement and chiral symmetry breaking happen below the dark QCD scale $\Lambda_d$ with $N_f^2 - 1$ dark mesons in the low-energy theory.~\footnote{This counting of dark mesons works for $N_d \geq 3$. For $N_d=2$, the chiral symmetry breaking is $\SU{2 N_f} \rightarrow \mathrm{SP}(2 N_f)$ with $2 N_f^2 - N_f -1$ dark mesons~\cite{Peskin:1980gc}.} Different from the SM QCD, where the phase transition is a crossover one~\cite{Fodor:2001pe}, there is a wide range of model parameter space for the dark QCD phase transition to be first order. 

%--------------------------------------------
% The Model
%--------------------------------------------
\subsection*{The Model}

%=========
Let $\psi_i(x)$ for $i \in\{ 1, 2, \cdots, N_f \}$ be a collection of Dirac spinor fields or dark quark, and let $G_\mu^a(x)$ for $a \in \{ 1, 2, \cdots, N_d^2-1 \}$ be the dark gluon fields and a collection of real vector fields that form the connection of an $\SU{N_d}$ gauge group under which the $\psi_i$ transform in the fundamental representation.  
The properties of these particles and their interactions are given by the following Lagrangian 
\begin{align}\label{eq:Lagrangian}
	\Lscr = \sum_{i=1}^{N_f} \Bigl[ \bar{\psi}_i i \gamma^\mu D_\mu \psi_i - m_i \bar{\psi}_i \psi_i \Bigr] - \frac{1}{4} G_{\mu\nu}^a G^{\mu\nu\,a} - \frac{1}{4} \frac{\theta_d}{2\pi} \frac{g_d^2}{4\pi} G_{\mu\nu}^a \widetilde{G}^{\mu\nu \, a} 
	\com
\end{align}
where 
\begin{align}
	D_\mu \psi_i & = \partial_\mu \psi_i - i g_d G_\mu^a T^a \psi_i \,, \quad
	G_{\mu\nu}^a  = \partial_\mu G^a_\nu - \partial_\nu G^a_\mu + g_d f^{abc} G_\mu^b G_\nu^c  \,, \quad
	\widetilde{G}^{\mu\nu\,a}  = \frac{1}{2} \epsilon^{\mu\nu\rho\sigma} G_{\rho\sigma}^a 
	\per
\end{align}
The generators of $\SU{N_d}$ are denoted as $T^a$, and the structure constants are denoted by $f^{abc}$.  

%=========
The model parameters are the number of colors $N_d \in \{ 2, 3, 4, \cdots \}$, the number of flavors $N_f \in \{ 1, 2, 3, \cdots \}$, the dark gauge coupling $g_d \in [0, \infty)$, the mass parameters $m_i \in [0,\infty)$, and the theta parameter $\theta_d \in [0,2\pi)$.  
We will consider both the case of massless quarks, $m_i = 0$, and massive quarks, $m_i \neq 0$.  For simplicity, we assume that the model is $\mathcal{CP}$-conserving with $\theta_d = 0$. There could exist non-renormalizable operators for the SM sector interacting with the dark QCD sector, which will be introduced and discussed in a later section. 

%=========
The fermion mass term in \eref{eq:Lagrangian} can be written more generally as $m_{ij} \bar{\psi}_i \psi_j$ for $m_{ij} \in \Cbb$, but we have performed a field redefinition to write it as $m_i \bar{\psi}_i \psi_i$ with $m_i$ being real and nonnegative.  
For $m_i = 0$ the theory respects a chiral flavor symmetry, $\SU{N_f}_V \times \U{1}_V \times \SU{N_f}_A \times \U{1}_A$.  
The symmetry group $\U{1}_V$ has an associated conserved charge, which is the dark baryon number, $\U{1}_\Bd$; the dark gluons, dark quarks, and dark antiquarks have charges $Q_\Bd(G_a) = 0$, $Q_\Bd(\psi_i) = 1/N_d$, and $Q_\Bd(\bar{\psi}_i) = -1/N_d$, respectively.  
The axial $U(1)_A$ symmetry is anomalous under the dark QCD gauge interactions and does not lead to a light Nambu-Goldstone boson after spontaneous chiral symmetry breaking.  
For $m_i \neq 0$ the subgroup $\SU{N_f}_A \times \U{1}_A$ is explicitly broken.  

%--------------------------------------------
% Color confinement
%--------------------------------------------
\subsection*{Color confinement}

%=========
Quantum effects lead to the renormalization group (RG) flow of the coupling $g_d$.  
Let $\hat{g}_d(\mu)$ be the running coupling, and let $\mu$ be the renormalization scale.  
The RG flow equation is
\begin{align}\label{eq:beta_func}
	\mu \frac{d\hat{g}_d}{d\mu} = \beta_{g_d} = \frac{\hat{g}_d^3}{16 \pi^2} b_{g_d} + O(\hat{g}_d^5)
	\com
\end{align}
and the leading-order term given by~\cite{Gross:1973id,Politzer:1973fx}
\begin{align}
	b_{g_d} = - \frac{11}{3} N_d + \frac{2}{3} N_f 
	\com
\end{align}
which can be negative.  

%=========
We are interested in models with $N_f < 11N_d/2$ for which $b_{g_d} < 0$, and the theory becomes more strongly coupled in the IR (smaller $\mu$).  
If we take $\hat{g}_d(\mu_\UV) = g_\UV$ as a reference point where the theory is weakly coupled, $g_\UV \ll 4\pi$, then by solving the RG flow equation we observe that $\hat{g}_d(\mu)$ diverges at $\mu = \mu_\ast$.  
As the gauge coupling becomes larger, the interactions among quarks and gluons become stronger, leading to a color-confining/chiral-symmetry-breaking phase of the theory.  
The value of $\mu_\ast$ provides a rough (one-loop perturbative) estimate of the confinement scale, $\Lambda_d \approx \mu_\ast$, which gives 
\begin{align} \label{eq:Lambda-d}
	\Lambda_d \approx \mu_\UV \ \mathrm{exp}\bigl[- 8 \pi^2 \, / \, (|b_{g_d}| \, g_\UV^2) \bigr] 
	\com
\end{align}
assuming that $b_{g_d} < 0$.  

%=========
Around the confinement scale, the fermion-anti-fermion operator also develops a nonzero expectation value with $\langle \overline{\psi}\psi \rangle \sim \Lambda_d^3$, which spontaneously breaks the $\SU{N_f}_A$ flavor symmetry and provides dark mesons as IR degrees of freedom.  
The dark meson decay constant is $f_{\pi_d} \sim \Lambda_d$, while their masses are related to the dark quark masses by $m_{\pi_d}^2 f^2_{\pi_d} \sim m_i \, \Lambda_d^3$.  
The dark baryon masses have $m_\dB \sim 4\pi \,\Lambda_d$ and are heavier.  
The temperature of the confining/chiral-symmetry-breaking phase transition happens at $T_c \sim \Lambda_d$.    
Some of our later calculations will be sensitive to some ratios of quantities like $m_\dB / T_c$, which requires a non-perturbative tool like lattice QCD to obtain a precise value.  

%--------------------------------------------
% Confining phase transition
%--------------------------------------------
\subsection*{Confining phase transition}

%=========
Let us now consider the behavior of this theory in a finite-temperature system, and specifically we are interested in a system whose temperature is close to the critical temperature of the confining phase transition, $T \sim T_c$.  
The order of magnitude of the critical temperature is set by the confinement scale, $T_c \sim \Lambda_d$.
Suppose that the system is heated to a temperature $T > T_c$ and allowed to cool adiabatically to $T \sim T_c$.  
Since the temperature sets the typical momentum transfer $|\Delta \pvec|$ of particles in the plasma, the system will be in the unconfined phase for $T > T_c \sim \Lambda_d$ where $|\Delta \pvec| \sim T > \Lambda_d$.  
However, as the temperature reaches close to $\Lambda_d$ the system will pass into the confined phase.  
At the same time a chiral condensate forms, $\langle \bar{\psi} \psi \rangle \neq 0$, signaling that the chiral symmetry is spontaneously broken.~\footnote{We assume that both chiral symmetry breaking and color confinement occur at around the same time during the phase transition at $T = T_c \sim\Lambda_d$.
}  
We are interested in whether the corresponding phase transition is a first order one, which is one of the necessary conditions to form the dark quark nuggets.

%=========
The order of this phase transition has been studied on general grounds by Pisarski and Wilczek (PW)~\cite{Pisarski:1983ms} for $N_d \geq 3$ (see Ref.~\cite{Wirstam:1999ds} for the $N_d=2$ case).  
Using a perturbative $\epsilon$-expansion, they argue that the chiral phase transition will be first order if the number of light vector-like fermion flavors is greater than or equal to three; in our notation, this corresponds to 
\begin{align}
	\text{PW argument: \qquad $N_f \geq 3$ \ for \ $m_i \ll \Lambda_d$  \ $\Rightarrow$ \  $1^\mathrm{st}$ order phase transition} 
	\per
\end{align} 
The essence of the argument is to write down an effective field theory describing the self-interactions of the chiral condensate, $\Sigma_{ij} \sim \langle \bar{\psi}_i (1 + \gamma_5) \psi_j \rangle$ with $i, j = 1, 2, \cdots, N_f$. Besides the instanton-generated $U(1)_A$-breaking term that is suppressed in the large $N_d$ limit, there are two couplings associated with the self-interaction operators, $( \mathrm{Tr} \, \Sigma^\dagger \Sigma )^2$ and $\mathrm{Tr} (\Sigma^\dagger \Sigma)^2$.  
PW calculate the beta functions for these couplings and argue that for $N_f \geq 3$ the RG flow equations do not have an IR stable fixed point.  
In the absence of an IR stable fixed point, the theory cannot be smoothly evolved to arbitrarily low scales (temperatures), but instead some critical behavior must arise in the form of a first order phase transition.  

%=========
Whereas the PW argument infers the existence of a first order phase transition indirectly from RG flow trajectories in the chiral effective theory, one can also study the phase transition directly by evaluating the thermal effective potential for the chiral condensate and calculating the thermal transition rate between coexistent phases.  
To justify a perturbative calculation of the effective potential, this approach is only reliable when the couplings are small, but nevertheless we can infer the behavior at a strong coupling by studying the trending behavior as the coupling is increased toward the non-perturbative regime.  
The results of this analysis are detailed in \aref{app:EFT}; in particular, we confirm that the chiral effective theory admits a first order phase transition in the regime consistent with the PW argument.  

%=============
\begin{figure}[th!]
\begin{center}
\includegraphics[width=0.5\textwidth]{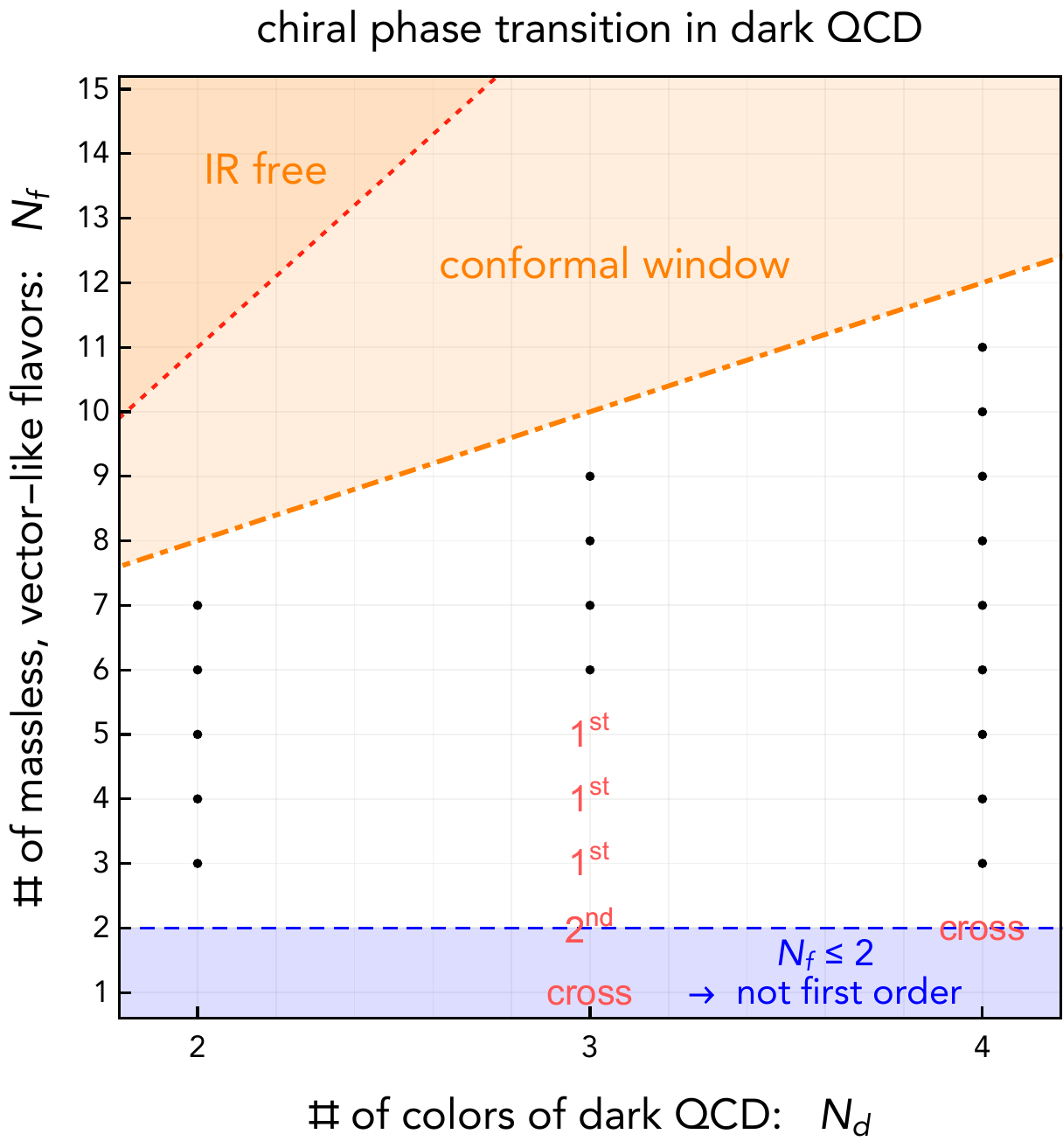} 
\caption{\label{fig:Nc_Nf}
The nature of the chiral phase transition in dark QCD is controlled by the number of colors, $N_d$, and the number of massless, vector-like flavors of fermions, $N_f$.  Points labeled by $1^\mathrm{st}$, $2^\mathrm{nd}$, and ``$\mathrm{cross}$" are known from lattice studies~\cite{Alexandrou:1998wv,Basile:2005hw,Lucini:2012gg,Brandt:2016daq,Ayyar:2018ppa} to exhibit a first order phase transition, a second order phase transition, and a continuous crossover, respectively.  Analytical arguments~\cite{Pisarski:1983ms,Wirstam:1999ds} imply that points falling into the unshaded (white) region will exhibit a first order phase transition.  The theory is not confining in the orange shaded regions:  above the dotted line the beta function remains positive, and between the dotted and dot-dashed lines, the theory becomes conformal at low energies.  The precise location of the conformal window's boundaries is a matter of active debate~\cite{DeGrand:2015zxa}.
}
\end{center}
\end{figure}

%=========
Since the PW argument is inherently perturbative in nature, one might worry that its conclusions do not apply for a strongly-coupled system.  
Thus it is important to ``test'' the PW argument against numerical lattice studies of the chiral phase transition.  
In \fref{fig:Nc_Nf} we summarize the results of several lattice studies for different values of $N_d$ and $N_f$ (assuming massless quarks/antiquarks) for $N_d=3, 4$~\cite{Alexandrou:1998wv,Basile:2005hw,Lucini:2012gg,Brandt:2016daq,Ayyar:2018ppa}.  We conclude that the PW argument is supported by numerical lattice simulations, which take all non-perturbative effects into account.  For $N_d=2$, more lattice QCD simulations are required to determine the order of phase transition~\cite{Hands:2010gd,Braguta:2014ira}. 

%=========
In \fref{fig:Nc_Nf}, we also indicate the parameter region where the leading-order beta-function is positive and the theory is ``IR-free'' rather than exhibiting confinement or chiral symmetry breaking at low energies.  
For smaller values of $N_f$, the ``conformal window'' corresponds to a range of parameters in which the theory goes to a nontrivial fixed point in the IR, and there is neither confinement nor chiral symmetry breaking.  
The boundary between the conformal window and models with chiral symmetry breaking (at smaller $N_f$) is an active subject of research for both lattice QCD or other semi-analytic approaches.  
In our plot, we take the point of view based on the review paper in \rref{DeGrand:2015zxa}: the conformal window line is determined by $N_d=2$ and $N_f \gtrsim 8$~\cite{Appelquist:2013pqa,Leino:2017lpc} and $N_d=3$ and $N_f \gtrsim 10$~\cite{Hayakawa:2010yn,Appelquist:2014zsa}. In the dotdashed line of \fref{fig:Nc_Nf}, we simply use the information at $N_d=2,3$ to obtain the conformal window boundary line as $N_f \approx 2 N_d + 4$. 

%=========
Finally let us remark on the range of interest for the model parameters.  
We will take $N_f \geq 3$ to ensure a first order chiral phase transition, and we will take $N_f \lesssim  2 N_d + 4$,  to ensure that confinement occurs.  Then the parameter range of interest is 
\begin{align}\label{eq:Nf_range}
	3 \leq N_f \lesssim 2 N_d + 4
	\qquad \text{with} \qquad 
	m_i \ll \Lambda_d 
	\per
\end{align}
We want to stress that there is a wide range of parameter space in $(N_d, N_f)$ for the dark QCD phase transition to be a first-order one. 

%--------------------------------------------
% Differential vacuum pressure:  $B$
%--------------------------------------------
\subsection*{Differential vacuum pressure:  $B$}

%=========
During the confining/chiral-symmetry phase transition, the system passes from a phase in which color is unconfined and the chiral symmetry is unbroken into a second phase in which color is confined and the chiral symmetry is broken.  
In general the vacuum energy of these two phases will differ, and it is the lower vacuum energy of the confined phase that makes the phase transition energetically favorable at low temperature.  
Since the vacuum has an equation of state, $\rho = - P$, we can equally well talk about the differential vacuum pressure between the two phases.  
Following the notation of the MIT bag model of SM nuclear structure~\cite{Hasenfratz:1977dt}, we denote this differential vacuum pressure as $B$, which has mass dimension equal to $4$.    
In principle $B$ can be expressed in terms of the model parameters: $\Lambda_d$, $N_d$, $N_f$, and $m_i$.  
However, a robust calculation of $B$ requires non-perturbative methods, such as numerical lattice techniques.  
Therefore we will generally take $B$ as a free parameter, while keeping in mind that it is roughly set by the confinement scale:  
\begin{align}\label{eq:B_def}
	B = \Delta P_\mathrm{vacuum} = P_\mathrm{confined} - P_\mathrm{unconfined} \sim \Lambda_d^4 
	\per
\end{align}
In \sref{sec:dark_quark_matter} we will see that $B$ controls the density and energy of the dark quark matter that resides inside of dark quark nuggets.  
Consequently in \sref{sec:nuggets} we will find that $B$ also sets the mass scale and radius of cosmologically-produced dark quark nuggets.  

%==================================
% Dark quark matter
%==================================
\section{Dark quark matter}\label{sec:dark_quark_matter}

%=========
The theory discussed in \sref{sec:dark_QCD} admits a state of ``dark quark matter'' (dQM) at zero temperature and finite dark-baryon-number density.  
In this section we calculate the thermodynamic properties of dQM by adapting a similar calculation from \rref{Witten:1984rs}.  
The main results of this section appear in \erefs{eq:rho_dQM}{eq:n_Bd_dQM}, which give energy density and the dark-baryon-number density of the dark quark matter contained within a stable dark quark nugget.  

%--------------------------------------------
% Modeling dQM as a relativistic degenerate Fermi gas
%--------------------------------------------
\subsection*{Modeling dQM as a relativistic degenerate Fermi gas}

%=========
We suppose that the model from \sref{sec:dark_QCD} is brought to a finite temperature $T$ where the dark gluons, dark quarks, and dark antiquarks are allowed to reach thermal equilibrium.  
We further suppose that the system is prepared with a nonzero dark baryon number.  
%Then the phase space distribution functions for the dark gluons, dark quarks, and dark antiquarks can be written as 
%\begin{align}
%	f_{G_a}(\pvec) & = \bigl\{ e^{[E_{G_a}(\pvec) - \mu_{G_a}]/T} - 1 \bigr\}^{-1} 
%	&& \text{for} 
%	& a & = 1, 2 ,\ldots , N_d^2 - 1  \,, \\ 
%	f_{\psi_i}(\pvec) & = \bigl\{ e^{[E_{\psi_i}(\pvec) - \mu_{\psi_i}]/T} + 1 \bigr\}^{-1} 
%	&& \text{for} 
%	& i & = 1, 2 ,\ldots , N_f  \,, \\ 
%	f_{\bar{\psi}_i}(\pvec) & = \bigl\{ e^{[E_{\bar{\psi}_i}(\pvec) - \mu_{\bar{\psi}_i}]/T} + 1 \bigr\}^{-1} 
%	&& \text{for} 
%	& i & = 1, 2 ,\ldots , N_f  \,,
%\end{align}
%where the $E$'s and $\mu$'s are the energies and chemical potentials of the corresponding particle.  

%=========
Dark QCD mediates interactions among the dark gluons and the dark quarks/antiquarks.  
If reactions such as $\psi_i \bar{\psi}_i \leftrightarrow G_a G_b$ and $\psi_i \bar{\psi}_i \leftrightarrow G_a G_b G_c$ are in thermal equilibrium, {\it i.e.} the thermally-averaged rate exceeds the Hubble expansion rate at the time of interest, then chemical equilibrium imposes $\mu_{G_a} = 0$ and $\mu_{\bar{\psi}_i} = - \mu_{\psi_i}$, where $\mu$ is the chemical potential of the species.  
%This leaves $N_f$ undetermined chemical potentials, which are related to $N_f$ conservation laws ($\psi_i$-number for $i = 1, 2, \ldots, N_f$).  
For simplicity, we further suppose that dark baryon number is shared equally by all of the quark and antiquark flavors, which implies that the chemical potentials are equal, $\mu_{\psi_i} = \mu$, and we also assume that the $N_f$ flavors of dark quarks and antiquarks are degenerate, which lets us write $m_i = m$; these assumptions does not qualitatively impact our results.  

%=========
We are interested in this system at a temperature $m_i \ll T \ll \mu$ such that the quarks and antiquarks form a relativistic degenerate Fermi gas~\cite{Huang_1987}.  
Let $n = n_{\psi} - n_{\bar{\psi}}$ be the $\psi$-number density, which contains an implicit sum over the $N_f$ flavors; let $\rho = \rho_{\psi} + \rho_{\bar{\psi}} + \rho_\mathrm{vacuum}$ be the energy density of quarks, antiquarks, and the dark quark matter vacuum; and let $P = P_{\psi} + P_{\bar{\psi}} + P_\mathrm{vacuum}$ be the corresponding pressure.  
For a relativistic degenerate Fermi gas, and neglecting the perturbative interactions among dark quarks and gluons, these quantities are given by~\cite{Huang_1987}
\begin{align}\label{eq:n_rho_P}
	n = g \, \frac{\mu^3}{6\pi^2} 
	\ , \qquad 
	\rho = g \, \frac{\mu^4}{8\pi^2} + B 
	\ , \qquad \text{and} \qquad 
	P = g \, \frac{\mu^4}{24\pi^2} - B
	\com
\end{align}
where $B$ is the differential vacuum pressure from \eref{eq:B_def} (normalized such that pressure vanishes in the hadronic phase) and where $g = 2 N_d N_f$ accounts for a sum over identically-distributed particles that differ in their spin, color, and flavor.  
The number density of dark baryon number is given by 
\begin{align}\label{eq:baryon_number_density}
	n_\Bd = \frac{1}{N_d} n = N_f \, \frac{\mu^3}{3\pi^2} 
	\com
\end{align}
since each dark quark carries a baryon number of $1/N_d$ and each antiquark has $-1/N_d$.  
Note that $n_\Bd$ is independent of $N_d$; raising $N_d$ means that there are more species of dark quarks/antiquarks in the system, but that each one carries a smaller dark baryon number.  

%=============
\begin{figure}[h!]
\begin{center}
\includegraphics[width=0.95\textwidth]{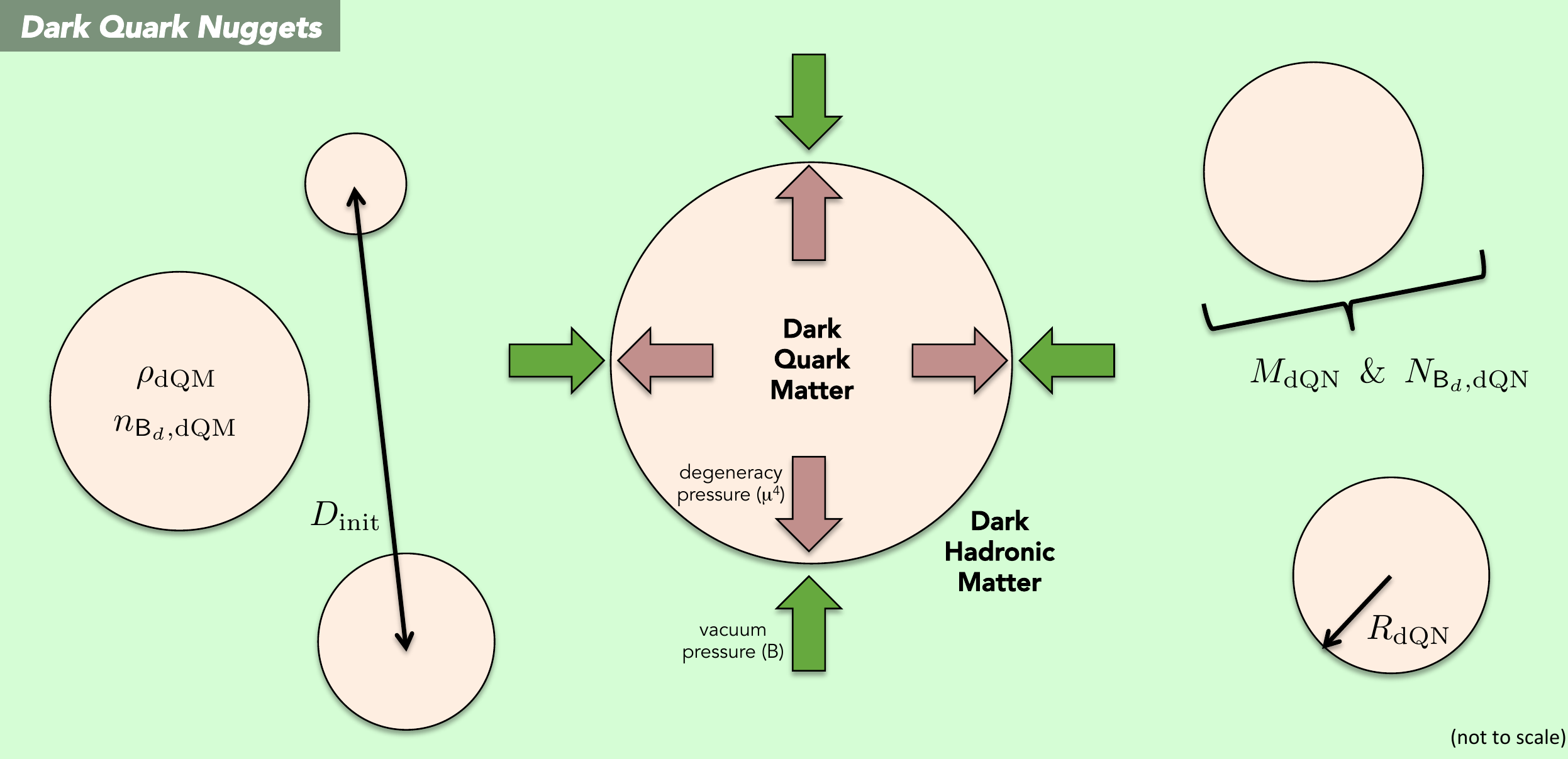} 
\caption{\label{fig:cartoon}
This cartoon illustrates the localized nugget of dark quark matter, which is supported against collapse by the Fermi degeneracy pressure arising from its conserved dark baryon number. 
}
\end{center}
\end{figure}

%--------------------------------------------
% Dark quark matter inside of nuggets
%--------------------------------------------
\subsection*{Dark quark matter inside of nuggets}

%=========
Now we suppose that the conserved dark baryon number is localized in a region of space with finite volume.  
If the volume is allowed to vary, such as during the formation of a dark quark nugget, then the system will evolve to an equilibrium configuration in which the differential vacuum pressure at the phase boundary is balanced against the differential pressure arising from the particles, $\Delta P_\mathrm{vacuum} = \Delta P_\mathrm{particles}$~\footnote{The gravitational pressure is negligible for the range of dQN masses considered in this paper. }.  
Here we assume that the plasma temperature is small compared to the phase transition temperature, which lets us write $\Delta P_\mathrm{vacuum} \approx B$ where $B$ is the differential vacuum pressure at zero temperature.  
We also continue to assume that $T \ll \mu$, which lets us neglect the radiation pressure that would arise from particles outside of the nugget and instead write $\Delta P_\mathrm{particles} \approx g \mu^4 / 24\pi^2$.
A cartoon of this situation is illustrated in \fref{fig:cartoon}.  
Thus the equilibrium condition is expressed as 
\begin{align}\label{eq:pressure_balance}
	P \bigr|_{\mu = \mu_\mathrm{eq}} = g \, \frac{\mu_\mathrm{eq}^4}{24 \pi^2} - B = 0
	\com
\end{align}
and its solution is $\mu_\mathrm{eq} \approx \bigl[ 12 \pi^2/(N_d N_f) \bigr]^{1/4} \, B^{1/4}$.  
For instance $N_d = N_f = 3$ gives $\mu_\mathrm{eq}  \simeq 1.9 \, B^{1/4}$.  

%=========
Now we are equipped to calculate the properties of the dark quark matter that resides inside of a stable dark quark nugget.  
The energy density of the dark quark matter inside of a dark quark nugget is calculated using $\rho$ from \eref{eq:n_rho_P} and $\mu = \mu_\mathrm{eq}$ from \eref{eq:pressure_balance}, which gives 
\begin{align}\label{eq:rho_dQM}
	\rho_\dQM = 4B
	\com
\end{align}
and the density of dark baryon number is evaluated with $n_\Bd$ from \eref{eq:baryon_number_density}, which gives 
\begin{align}\label{eq:n_Bd_dQM}
	n_{\Bd,\dQM} = \left( \frac{64 N_f}{3 \pi^2 N_d^3} \right)^{1/4} B^{3/4} 
	\per
\end{align}
Thus the energy per baryon of dark quark matter in dark quark nuggets is found to be 
\begin{align}\label{eq:energy_per_baryon}
	\frac{\rho_\dQM}{n_{\Bd,\dQM}} = \left( \frac{12 \pi^2 N_d^3}{N_f} \right)^{1/4} B^{1/4} \simeq 3.3 \ \frac{N_d^{3/4}}{N_f^{1/4}} \ B^{1/4}
	\per 
\end{align}
For instance $N_d = N_f = 3$ gives $5.7 \, B^{1/4}$.

%=========
Looking back over these results, we observe that the differential vacuum pressure between the confined and unconfined phases, $\Delta P_\mathrm{vacuum} = B$ from \eref{eq:B_def}, is the only scale that sets the density and energy of the dark quark matter that resides inside of dark quark nuggets.  
We will use \erefs{eq:rho_dQM}{eq:n_Bd_dQM} in \sref{sec:nuggets} to estimate the size and mass of a typical dark quark nugget, and we will use \eref{eq:energy_per_baryon} in the subsection below to discuss stability of dark quark matter.  

%--------------------------------------------
% Stability of dark quark matter
%--------------------------------------------
\subsection*{Stability of dark quark matter}

%=========
The quantity $\rho_\dQM/n_{\Bd,\dQM}$ is used to assess whether the state of dark quark matter is more or less stable than the state of dark hadronic matter.  
Suppose that the lightest stable dark baryons are all degenerate and let their mass be denoted by $m_\dB$.  
In the dark hadronic state and for a volume of $V$, a state with $n_\Bd V$ units of dark baryon number can have an energy that is as low as $m_\dB\,n_\Bd\,V$ (if all the dark baryons are at rest with negligible interactions and no additional particles are present).  
Thus the state of dark quark matter is absolutely stable provided that $\rho \, V < m_\dB\,n_\Bd\,V$.  
Using the expression for $\rho/n_\Bd$ from \eref{eq:energy_per_baryon}, the stability of dark quark matter requires 
\begin{align} \label{eq:stability-condition}
	\frac{B^{1/4}}{m_\dB} < 
	0.175 \, \left( \frac{N_f/N_d}{1} \right)^{1/4} \left( \frac{N_d}{3} \right)^{-1/2}
	\per
\end{align}
Recall that we need $N_f/N_d \gtrsim 1$ for a first order phase transition.  
Both the differential vacuum energy, $B$, and the dark baryon mass, $m_\dB$, are controlled by the confinement scale of the dark QCD, $\Lambda_d$.  
In SM QCD we have $B^{1/4} \simeq 150 \MeV$ and $m_\dB \simeq 938 \MeV$ to give $B^{1/4} / m_\dB \simeq 0.160$~\cite{Farhi:1984qu}.  
For a generic dark QCD model, a non-perturbative tool like lattice QCD is needed to estimate this ratio precisely.  
For a fixed value of $N_d$, there is a critical value of the number of flavors, $N_f = N_f^c$, above which the infrared theory of dark QCD becomes conformal instead of chiral symmetry breaking.  
When the number of flavor is close to the critical value, we anticipate that this ratio is further suppressed and scales like $B^{1/4} / m_\dB \propto (N_f^c - N_f)/N_f$~\cite{Bai:2018vik}. So, the dark quark matter state becomes more stable for a larger value of $N_f$.  

%=========
In \eref{eq:stability-condition}, we have only compared the quark matter state with a free baryon state.  
In the SM QCD, the most stable state per baryon is the iron nucleus, which has the energy per baryon slightly smaller ($\approx 1\%$) than a free proton and neutron.  
So, if the value of $B^{1/4} / m_\dB$ is so close to the upper bound in \eref{eq:stability-condition}, one may need to check the additional heavy-dark-nuclei evaporation processes, which will depend on more detailed properties of the model like the dark-meson-induced binding energy. For the massless dark meson case or the chiral limit, the inter-nucleon binding energy is anticipated to be larger by only a factor of around 2 than the SM QCD case~\cite{Berengut:2013nh}, so for a wide range of model parameters not saturating the bound in \eref{eq:stability-condition}, one does not need to worry about evaporation to heavy dark nuclei.     

%=========
%Other than checking the stability of dark quark nuggets during the current universe with a low temperature, one may also worry about its evaporation at a temperature not that far below the phase transition temperature. 
Similar to the SM QCD nugget scenario, the equilibrium between the two phases at temperature below $T_c$ is maintained by surface evaporation and emission of light particles. The detailed calculation on the establishment of the equilibrium is complicated. Here we would only provide simple pictures and argue that the nuggets may survive the evaporation and meanwhile stay thermalized with the plasma. In surface evaporation, the nugget emits a dark baryon and undergoes $(N_\B + 1) \rightarrow N_\B +1$~\cite{Alcock:1985vc}. However, such processes require addition energy input from the environment, as argued above. In SM QCD the energy is dumped into the nuggets by neutrinos, which has a long free-streaming length of $\mathcal{O}(0.1~\mbox{m})$ at the QCD scale. As we have no dark neutrinos in our model, the energy carrier in the dark quark nugget scenario will be the massless dark pions.
%For the SM 3-flavor quark matter, the neutrinos have a long free-streaming length. With enough energy, it can kick out a neutron from the quark matter and induce the surface evaporation process for the quark matter $(N_\B + 1) \rightarrow N_\B +1$~\cite{Alcock:1985vc} (see also Ref.~\cite{Madsen:1986jg} for the important reabsorption effects). 
%For the dark quark nuggets, the dark mesons in the dark hadron plasma can carry enough energy to evaporate the dark quark nuggets. 
However, because the strong interactions of dark mesons with other hadrons, their free-streaming length is very short at the order of $10^2/{T_c}$ and around $100\,\mbox{fm}$ for $T_c=0.1\,\mbox{GeV}$ and $m_{\dB}/T_c \approx 7$.
% (the neutrino free-streaming length at this temperature is around one meter and much longer). 
This much shorter length compared to the neutrino one can lead a dramatical reduction on the energy injection and hence the evaporation rate, and make the dark quark nugget more stable against the evaporation process. In addition, it has been argued that reabsorption effect will further enhance the stability of nuggets against evaporation~\cite{Madsen:1986jg}. Therefore, we would ignore the dark baryon dissipation from evaporation in the following analysis.

Since there is no dark neutrino in our model, one may wonder whether the dark quark nuggets will stay ``hot" after their formation below the phase transition. We want to point it out that the dark mesons can efficiently thermalize the dark quark nuggets with the surrounding medium and make nuggets cool as the universe cools down. Because dark mesons have a short free-streaming length, the cooling of nuggets is mainly through surface evaporation of dark mesons from black-body radiation. To simplify our discussion, we keep the chemical potential and radius of the nuggets fixed, which is reasonable within a Hubble time scale. We will check the cooling time scales for both an earlier time with a tiny chemical potential and a later time with a large chemical potential. 

Using the Stefan-Boltzmann law of black-body radiation, we have the cooling rate given by
\begin{align}
L(T)=\frac{\pi^3}{30}\,g_{\rm d\pi}\,R^2\,T^4,\qquad \mbox{with} \quad g_{\rm d\pi}=N^2_f-1  \,.
\end{align}
The total energy inside has $E(T)=\frac{4}{3}\pi R^3 \rho(T)$ with $\rho= g_{\rm dQ}\,\pi^2\,T^4/30$ when $\mu \ll T$ and $\rho=g_{\rm dQ}\, (\mu^4 + 2 \pi^2\, \mu^2 \, T^2)/8\pi^2$ for $T \ll \mu$. Here, we take the degrees of freedom as $g_{\rm dQ}=(2\,N_d\,N_f)\times 7/8$ for a temperature after the dark quark and dark anti-quark annihilation. Using the energy conservation $dE(T)/dt = - L(T)$, we can derive a differential equation for the temperature change as a function time and have the cooling time scale (the time for the temperature decreases from $T$ to $T/2$) estimated as
\begin{align}
\tau_{\rm cool}=
	\begin{cases}
	\dfrac{16 \ln2}{3}\dfrac{g_{\rm dQ}}{g_{\rm d\pi}}R   \,, & \mu \ll T \, ; \vspace{2mm}\\
	\dfrac{10}{\pi^2}\dfrac{\mu^2}{T^2}\dfrac{g_{\rm dQ}}{g_{\rm d\pi}}\,R  \,, & \mu \gg T \,.\\
	\end{cases}
\end{align}
When the temperature is high, the nugget radius is smaller than the Hubble scale $R \sim 10^{-5} \, d_H$ because there are around $10^{14}$ nucleation sits within one Hubble volume (see Appendix~\ref{app:EFT}). So the cooling time is shorter than the Hubble expansion time. When the chemical potential is high or temperature is low, one has $\tau_{\rm cool}/t_H \propto \mu^2 R / M_{\rm pl} \sim 10^{-8}$ for the benchmark point with $\mu \sim T_c =100$~MeV and $R \sim 0.1$~cm from \eref{eq:R_dQN_alt}. The cooling time has a mild $T_c$ dependence: $\tau_{\rm cool} \sim T_c^{-1/3}$, so we have a sufficiently fast thermalization for the nuggets with the surrounding medium for the model parameter space in this paper. 

\section{Cosmological production of dark quark nuggets}\label{sec:dark_nuggets}

%=========
In this section we discuss how dark quark nuggets can form in the early universe, we calculate their properties and estimate their relic abundance.  

%--------------------------------------------
% Overview of dark quark nugget production
%--------------------------------------------
\subsection{Overview of dark quark nugget production}\label{sec:Overview}

\begin{figure}[h!]
\begin{center}
\includegraphics[width=0.65\textwidth]{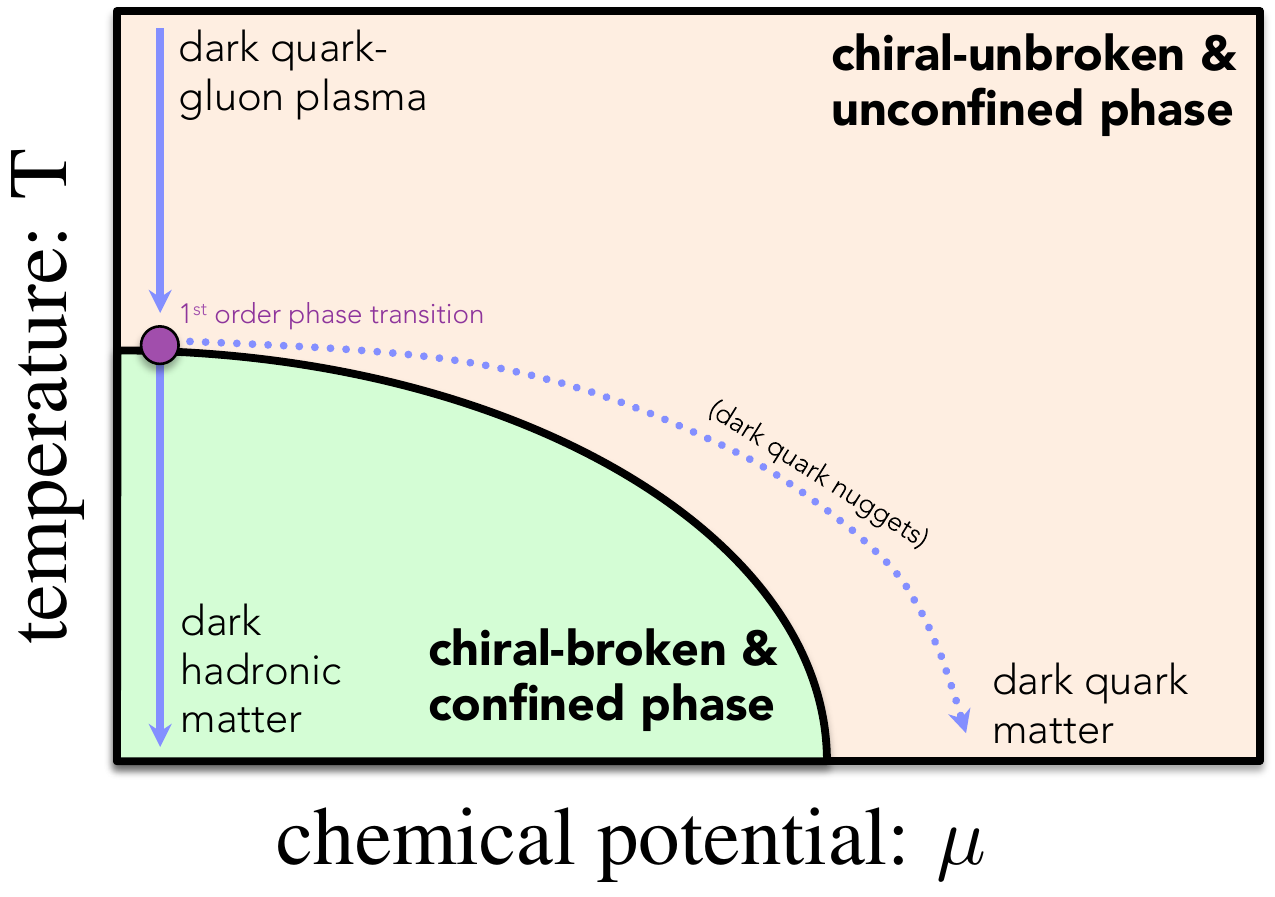} 
\caption{\label{fig:phase_diagram}
A schematic phase diagram of the dark QCD sector is shown here along with the trajectory through phase space during the formation of dark quark nuggets.  The entire system is initially in the unconfined phase at high temperature and small chemical potential (corresponding to the nonzero dark baryon asymmetry).  The system cools due to cosmological expansion, which triggers a first order phase transition.  Some regions of space enter the confined phase where the dark baryon asymmetry is eventually carried by free dark baryons and antibaryons, but most of the dark baryon asymmetry is collected into pockets of space that cool to form dark quark nuggets.  If the chemical potential is large, there may be exotic phases, similar to the color superconductivity and the color-flavor-locking phase of QCD~\cite{Alford:2007xm}, but we neglect this possibility.
}
\end{center}
\end{figure}

%=========
Dark quark nuggets may form at a first order phase transition during which dark color is confined and the chiral symmetry is spontaneously broken.
The production mechanism for dark quark nuggets is very similar to the more-familiar QCD quark nugget scenario~\cite{Witten:1984rs}.   
Here we briefly summarize the physical processes that lead to creation of dark quark nuggets in the early universe.  
The production process is also illustrated in \fref{fig:phase_diagram} that shows a schematic phase diagram for dark QCD.  
\begin{enumerate}
	\item The dark sector and the SM sector remain thermalized with each other until they decouple at a temperature $T_\mathrm{dec}$.  Afterward the temperatures of the two sectors evolve independently, decreasing with the adiabatic expansion of the universe.  
	\item As the temperature of the dark sector cools down to a temperature $T_\ast$ slightly below the critical temperature $T_c$, the bubbles of dark hadrons start to nucleate out of the dark quark-gluon plasma. The pressure difference $\Delta P=B$ between the two phases drives the growth of the bubbles, while the scattering of the particles in the dark plasma on the bubble wall induces a drag force on the bubble wall.  A balance between vacuum pressure and thermal pressure is reached and the bubble's radius grows at a nonrelativistic terminal speed.
	\item  It is energetically preferable for dark baryon number to remain in the unconfined phase, where dark quarks are light, rather than entering the confined phase, where dark baryons are heavy.  Thus, dark baryon number accumulates in front of the advancing bubble walls.  
	\item The bubbles collide and coalescence with each other.  At the end of the phase transition, the dark hadron phase occupies the majority of the Hubble volume, with the remaining dark quark-gluon plasma left in isolated regions that form dark quark nuggets. Most of the dark baryon number is stored in dQN with the remainder carried by free dark baryons.  
	\item After the phase transition, the cosmological plasma continues to cool and the remaining regions of dark quark-gluon plasma shrink as the thermal pressure decreases.  When the temperature decreases below the chemical potential in these regions, they become dark quark nuggets, supported by degeneracy Fermi pressure.  
\end{enumerate}

%=========

%--------------------------------------------
% Dark baryon number accumulates
%--------------------------------------------
\subsection{Dark baryon number accumulates in the quark nuggets}\label{sec:accumulates}

%=========
Particles in the plasma scatter from the passing bubble wall, and this causes dark baryon number to accumulate in the unbroken phase.  
In front of the wall, baryon number is carried by the dark quarks and antiquarks, which are approximately massless.  
However, behind the wall the dark baryon number is carried by dark baryons and antibaryons, which acquire a mass $m_\dB$.  
If $m_\dB$ is much larger than the temperature of the phase transition, $T_c$, then the amount of baryon number entering the bubble will be Boltzmann suppressed.  
%\SL{During this process, the massless dark radiation could serve as a messenger between the two phases, serving as the analogue of neutrinos in the formation of QCD nuggets.}

%=========
\rref{Bodeker:2009qy} has studied the kinematics of a particle scattering from a bubble wall where the particle's mass changes.  
By applying that analysis to the problem of dQN formation, we find that dark baryon number will be kinematically blocked from entering the confined-phase bubbles if the dark baryon mass is sufficiently large:  
\begin{align}\label{eq:mass_condit}
	m_\dB > 2 \, \gamma_w \, p_z \qquad \text{with} \qquad p_z \sim p_\mathrm{rms} \simeq 3.6 \, T_c
	\per
\end{align}
The factor of $3.6$ in the root-mean-square momentum follows from the Fermi-Dirac distribution.
Here $\gamma_w = 1 / \sqrt{1 - v_w^2}$ the wall's boost factor, and $v_w$ is its speed.  
It is challenging to calculate the wall's speed from first principles~\cite{Espinosa:2010hh}.  
(See also \rref{Kajantie:1992uk}, which estimates the maximum deflagration velocity allowed by entropy increase, and argues that $v_w$ is non-relativistic.)  
However, due to the strongly-coupled nature of the dark QCD interactions, we think it is reasonable to expect that particles in the plasma will induce a large drag force on the wall and lead to a non-relativistic terminal velocity with $\gamma_w \approx 1$.  
If that is the case, then \eref{eq:mass_condit} imposes a weak constraint, $m_\dB \gtrsim 7\,T_c$. For the model parameters satisfying this constraint, the dark baryon number is kinematically preferred to stay in the unbroken phase. Otherwise if $\gamma_w \gg 1$, effectively all particles in the plasma will have enough energy to enter the bubble, and the dark baryon number will hardly remain in the unbroken phase.%For the parameter region violating this constraint, the flow of the dark baryon number from one region to another region could be determined by the chemical and thermal equilibrium conditions, which will be discussed more later when we calculate the relic abundance of the dark quark nuggets. 

%--------------------------------------------
% Dark quark nuggets
%--------------------------------------------
\subsection{Dark quark nuggets: mass, size, and relic abundance}\label{sec:nuggets}

%=========
Let us now estimate the typical mass, size, and relic abundance of the dark quark nuggets.  
The notation used in this section is summarized in \tref{tab:notation}.  
Already in \sref{sec:dark_quark_matter} we have studied the dark quark matter that resides inside of a dark quark nugget, and we have calculated its energy density, $\rho_\dQM$, and number density of dark baryon number, $n_{\Bd,\dQM}$.  
Now all that remains is to estimate the typical amount of dark baryon number per nugget, $N_{\Bd,\dQN}$, and then the nugget's radius and mass are given by $(4\pi/3) \, R^3_\dQN \, n_{\Bd,\dQM} = N_{\Bd,\dQN}$ and $(4\pi/3) \, R^3_\dQN \, \rho_\dQM = M_\dQN$.  

%========= 
We assume that all the nuggets have a comparable amount of dark baryon number, and that this quantity is approximately conserved from the time of nugget formation until today.  
Thus we can write $N_{\Bd,\dQN} = f_\mathrm{nug} \, n_\Bd^\mathrm{Hub}(t_c) / n_\dQN(t_c)$ where $n_\Bd^\mathrm{Hub}(t_c)$ is the cosmological density of dark baryon number at the time of the phase transition, $n_\dQN(t_c)$ is the cosmological density of dark quark nuggets at the time of the phase transition, and $f_\mathrm{nug}$ is the fraction of dark baryon number that gets stored in the dark quark nuggets (leaving a fraction $f_\mathrm{free}=1-f_\mathrm{nug}$ to be stored in free dark baryons).  

%=========
\begin{table}[h!]
\begin{center}
\begin{tabular}{c|c|c}
	Symbol & Definition & Equation \\ \hline
	$M_\dQN$ & mass of a typical dark quark nugget & Eqs.~(\ref{eq:M_dQN},~\ref{eq:M_dQN_alt}) \\ 
	$R_\dQN$ & radius of a typical dark quark nugget & Eqs.~(\ref{eq:R_dQN},~\ref{eq:R_dQN_alt}) \\ 
	$N_{\Bd,\dQN}$ & amount of dark baryon number in a typical dark quark nugget & \eref{eq:N_Bd_dQN} \\ \hline 
	$n_\dQN(t)$ & cosmological number density of dark quark nuggets at time $t$ & \eref{eq:n_dQN_tc} \\ 
	$\Omega_\dQN h^2$ & cosmological relic abundance of dark quark nuggets today & \eref{eq:Omega_dQN} \\ 
	$D_\mathrm{init}$ & typical inter-nugget separation distance at the phase transition & \eref{eq:D_init} \\ \hline
	$n_{\Bd,\dQM}$ & density of dark baryon number of the dQM inside of a dQN & \eref{eq:n_Bd_dQM} \\ 
	$\rho_\dQM$ & energy density of the dQM inside of a dQN & \eref{eq:rho_dQM} \\ \hline 
	$Y_\Bd$ & cosmological yield of dark baryon number (conserved) & \\ 
	$n_\Bd^\mathrm{Hub}(t)$ & cosmological density of dark baryon number at time $t$ & \\ 
	$N_\mathrm{bary}$ & dimension of the quasi-degenerate dark baryon multiplet & \eref{eq:N_bary} \\ 
	$m_\dB$ & mass of the quasi-degenerate dark baryon multiplet & \\  
	$f_\mathrm{nug} = 1 - f_\mathrm{free}$ & fraction of dark baryon number stored in dark quark nuggets & \eref{eq:f_nug} \\ \hline
	$T_d(t)$ \& $T_\gamma(t)$ & temperature of the dark and visible sectors at time $t$ & \\ 
	$g_{\ast,d}(t) \approx g_{\ast S,d}(t)$ & effective number of relativistic dark-sector species at time $t$ & \\ 
	$g_{\ast,\gamma}(t) \approx g_{\ast S,\gamma}(t)$ & effective number of relativistic visible-sector species at time $t$ & \\ 
	$T_c = T_d(t_c)$ & temperature of the dark sector during the phase transition & \\ 
	$\Tc = T_\gamma(t_c)$ & temperature of the visible sector during the phase transition & \\ \hline
\end{tabular}
\end{center}
\caption{\label{tab:notation}Notation used in this section.  }
\end{table}%

%=========
The cosmological density of dark baryon number can be written as $n_\Bd^\mathrm{Hub} = Y_\Bd s$ where $Y_\Bd$ is the cosmological dark baryon number yield, and $s$ is the cosmological entropy density.  
We take the yield, $Y_\Bd$, as a free parameter and note for reference that the cosmological yield of SM baryon number is measured to be $Y_\mathsf{B} \simeq 10^{-10}$~\cite{Aghanim:2018eyx}. 
The entropy density can be written as $s=(2\pi^2/45) \, g_{\ast S} \, \Tc^3$ where $g_{\ast S} = g_{\ast S,\gamma} + g_{\ast S,d} \, \bigl( T_c / \Tc \bigr)^3$ counts the effective number of relativistic degrees of freedom in the plasma at the phase transition. Here, $\Tc$ is the temperature of the visible sector during the phase transition.   

%=========
We estimate the density of dark quark nuggets at the phase transition, $n_\dQN(t_c)$, by adopting the results of \aref{app:EFT}.  
In the appendix we study the dark QCD chiral phase transition using a chiral effective theory.  
The main result appears in \eref{eq:n_nucleations_numerical}, which gives $n_\mathrm{nucleations}$, the average number density of chiral-broken-phase bubbles that are nucleated over the course of the phase transition.  
We estimate that after the phase transition is completed, there is roughly one nugget produced for each nucleation, {\it i.e.} $n_\dQN(t_c) \approx n_\mathrm{nucleations}$.  
This lets us infer the density of dark quark nuggets at the end of the dark QCD phase transition to be 
\begin{align}\label{eq:n_dQN_tc}
	n_\dQN(t_c)
	\simeq \bigl( 2.1 \times 10^{14} \bigr) \left( \frac{\tilde{\sigma}}{0.1} \right)^{-9/2} H(t_c)^3
	\per
\end{align}
We have defined the dimensionless parameter $\tilde{\sigma} = \sigma / (B^{2/3} T_c^{1/3})$, and we have introduced $\sigma$, which represents the surface tension of a critical bubble at the time of nucleation; a larger value of $\sigma$ implies less efficient bubble nucleation, fewer nucleation sites, and more dark baryon number per nugget.  
The Hubble parameter is given by $3 \Mpl^2 H(t_c)^2 = (\pi^2/30) \, g_\ast(t_c) \, \Tc^4$ where $g_\ast(t_c) = g_{\ast,\gamma} + g_{\ast,d} \bigl[ T_d(t_c) / \Tc \bigr]^4$.  
The relation in \eref{eq:n_dQN_tc} reveals that there are typically $\sim 10^{14} \, (\tilde{\sigma}/0.1)^{-9/2}$ dark quark nuggets per Hubble volume, regardless of the temperature of the confining phase transition.  
The typical inter-nugget separation distance, $D_\mathrm{init}$, is then estimated as $D_\mathrm{init} = n_\dQN^{-1/3}$ to obtain 
\begin{align}\label{eq:D_init}
	D_\mathrm{init} 
	& \simeq \bigl( 77 \cm \bigr) \left[ \frac{g_\ast(t_c)}{10} \right]^{-1/2} \left( \frac{\Tc}{0.1 \GeV} \right)^{-2} \left( \frac{\tilde{\sigma}}{0.1} \right)^{3/2}
	\com
\end{align}
and for comparison the Hubble radius is $d_H \simeq 4.6 \times 10^6 \cm$.  

%=========
We estimate $f_\mathrm{nug}$ as follows.  
If the bubble wall expands sufficiently slowly, then thermal and chemical equilibrium is maintained at the phase boundary~\cite{Witten:1984rs}.  
It is energetically preferable for dark baryon number to remain in the unconfined phase where the dark quarks are massless, rather than enter the confined phase where the dark baryons acquire a mass $m_\dB \gg T_c$.  
From these considerations (for more details\footnote{Note that there is a typo in Eq.~(3.15) of the journal version of \rref{Bai:2018vik}; the value of $r$ is too large by a factor of $8$.  Upon correcting the error, the quark nugget relic abundance, $\Omega_\mathrm{QN} \sim 1/r$, is increased by a factor of $8$, and Fig.~5 of \rref{Bai:2018vik} is modified accordingly.  } see \rref{Bai:2018vik}) one can estimate the fraction of dark baryon number that goes into the dark quark nuggets to be 
\begin{align}\label{eq:f_nug}
	f_\mathrm{nug} 
	= 1- f_\mathrm{free} 
%	\approx \left[ 1 + \frac{N_\mathrm{bary} N_d}{N_f} \frac{\sqrt{2\pi}}{3 \zeta(3)} \, \left( \frac{m_\dB }{T_c} \right)^{3/2} e^{-m_\dB / T_c} \right]^{-1}
	\approx 1 - \frac{N_\mathrm{bary} N_d}{N_f} \frac{\sqrt{2\pi}}{3 \zeta(3)} \, \left( \frac{m_\dB }{T_c} \right)^{3/2} e^{-m_\dB / T_c} 
	\per
\end{align}
Here $N_\mathrm{bary}$ represents the number of quasi-degenerate baryons with mass $m_\dB$ in the confined phase (behind the bubble wall) for the lowest-spin and color-singlet state as a representation of the unbroken flavor symmetry $\SU{N_f}_V$.  
Using a simple group theory calculation,\footnote{These expressions are equal to the dimension of the representation of the baryon multiplet.  
The dimension is calculated with the aid of a Young tableau having two rows of $N_d/2$ boxes for even $N_d$, or two rows with $(N_d+1)/2$ and $(N_d-1)/2$ boxes for odd $N_d$~\cite{Antipin:2015xia}.  
For example, $N_\mathrm{bary} = 8$ for $N_d = N_f = 3$, reproducing the SM baryon octet.} one has
\begin{align}\label{eq:N_bary}
	N_\mathrm{bary} = 
	\begin{cases}
	\dfrac{\left(N_f + N_d / 2 - 1\right)!\, (N_f + N_d / 2 - 2)!}{(N_f - 1)!\, (N_f - 2)!\, (N_d / 2 + 1)!\, (N_d / 2)!} & , \quad \text{$N_d$ is even}\vspace{2mm}\\
	\dfrac{2\left(N_f + N_d / 2 - 1/2\right)!\, (N_f + N_d / 2 - 5/2)!}{(N_f - 1)!\, (N_f - 2)!\, (N_d / 2 + 3/2)!\, (N_d / 2 - 1/2)!} & , \quad \text{$N_d$ is odd}
	\end{cases} 
	\per
\end{align}
Taking $N_d = N_f = 3$ and $m_\dB/T_c=10$ gives $f_\mathrm{nug} \simeq 99.2\%$ and $f_\mathrm{free} \simeq 0.8\%$, meaning that most of the dark baryon number is stored in the dark quark nuggets.  

%=========
By combining the formulas for $n_\Bd^\mathrm{Hub}(t_c)$ and $n_\dQN(t_c)$, we estimate the amount of dark baryon number inside of a dark quark nugget to be 
\begin{align}\label{eq:N_Bd_dQN}
	N_{\Bd,\dQN} 
	\approx \frac{f_\mathrm{nug} \, n_\Bd^\mathrm{Hub}(t_c)}{n_\dQN(t_c)} 
	\simeq \left( 2.6 \times 10^{35} \right) 
	\left( \frac{f_\mathrm{nug}}{1} \right) 
	\left( \frac{Y_\Bd}{10^{-9}} \right) 
	\left(\frac{\Tc}{0.1 \GeV} \right)^{-3} 
	\left( \frac{\tilde{\sigma}}{0.1} \right)^{9/2}  \,,
\end{align}
where we have used $g_{\ast S} \approx g_\ast \simeq 10$. 
Here we have taken a fiducial value of $f_\mathrm{nug} = 1$, which corresponds to putting all of the dark baryon number into the dark quark nuggets (and leaving no dark baryon number for free dark baryons), but more generally the parameter $f_\mathrm{nug}$ can be related to the confinement scale and phase transition temperature through \eref{eq:f_nug}.  

%=========
Using the estimate for $N_\Bd$, it is now straightforward to estimate the radius and the mass of a typical dark quark nugget.  
The radius of the dark quark nugget satisfies $(4\pi/3) R^3_\dQN n_{\Bd,\dQM} = N_\Bd$ where the density of dark baryon number in the dark quark matter state is given by \eref{eq:n_Bd_dQM}.  
Solving for $R_\dQN$ gives the typical radius of a dark quark nugget to be 
\begin{align}\label{eq:R_dQN}
	R_\dQN 
	& \simeq \bigl( 0.073~\mathrm{cm} \bigr) 
	\left( \frac{N_d^{1/4}}{N_f^{1/12}} \right) 
	\left( \frac{B}{(0.1 \GeV)^4} \right)^{-1/4} 
	\left( \frac{f_\mathrm{nug}}{1} \right)^{1/3} 
	\left( \frac{Y_\Bd}{10^{-9}} \right)^{1/3} 
	\left( \frac{\Tc}{0.1 \GeV} \right)^{-1} 
	\left( \frac{\tilde{\sigma}}{0.1} \right)^{3/2} 
	\per
\end{align}
Similarly the mass of the dark quark nugget satisfies $(4\pi/3) R^3_\dQN \rho_\dQM = M_\dQN$ where the energy density of the dark quark matter is given by \eref{eq:rho_dQM}.  
This lets us estimate the typical nugget mass as 
\begin{align}\label{eq:M_dQN}
	M_\dQN
	& \simeq \bigl( 1.5 \times 10^{11} \gram \bigr) 
	\left( \frac{N_d^{3/4}}{N_f^{1/4}} \right) 
	\left( \frac{B}{(0.1 \GeV)^4} \right)^{1/4} 
	\left( \frac{f_\mathrm{nug}}{1} \right) 
	\left( \frac{Y_\Bd}{10^{-9}} \right) 
	\left( \frac{\Tc}{0.1 \GeV} \right)^{-3} 
	\left( \frac{\tilde{\sigma}}{0.1} \right)^{9/2}
	\per
\end{align}
Recall that $1 \times 10^{11} \gram \simeq 5 \times 10^{-23} \Msun$.  

%=========
Finally we estimate the relic abundance of dark quark nuggets in the universe today.  
Let $\Omega_\dQN = \rho_\dQN(t_0) / (3 \Mpl^2 H_0^2)$ where $\rho_\dQN(t_0)$ is the cosmological energy density of dark quark nuggets in the universe today and $H_0 = 100\,h \km / \mathrm{sec} / \mathrm{Mpc}$ with $h \simeq 0.674$~\cite{Aghanim:2018eyx}.  
Since the dark quark nuggets are nonrelativistic, we can write $\rho_\dQN(t_0) = M_\dQN \, n_\dQN(t_0)$ where $n_\dQN(t_0)$ is their cosmological number density today.  
If the nuggets do not merge or evaporate (see \sref{sec:cosmic_ray}) then their comoving number density, $n_\dQN(t) a(t)^3$, is conserved; here $a(t)$ is the Friedmann-Robertson-Walker (FRW) scale factor at time $t$.  
While the universe expands adiabatically, the comoving entropy density, $s(t) a(t)^3$, is conserved.  
Combining these formulas gives the relic abundance of dark quark nuggets today to be 
\begin{align}\label{eq:Omega_dQN}
	\Omega_\dQN h^2 
	& = \frac{M_\dQN \, n_\dQN(t_c)}{3 \Mpl^2 (H_0/h)^2} \,  
	\left( \frac{g_{\ast S}(t_0) \, T_\gamma(t_0)^3}{g_{\ast S}(t_c) \, \Tc^3} \right) 
\\
	& \simeq \bigl( 0.090 \bigr) 
	\left( \frac{N_d^{3/4}}{N_f^{1/4}} \right) 
	\left( \frac{B}{(0.1 \GeV)^4} \right)^{1/4} 
	\left( \frac{f_\mathrm{nug}}{1} \right) 
	\left( \frac{Y_\Bd}{10^{-9}} \right)
	\per \nonumber 
\end{align}
For reference, the relic abundance of dark matter is measured to be $\Omega_\DM h^2 \simeq 0.12$~\cite{Aghanim:2018eyx}.  
Thus the nuggets can make up all of the dark matter ($\Omega_\dQN h^2 \simeq 0.12$) if the differential vacuum pressure is at the nuclear energy scale, $B \simeq (0.1 \GeV)^4$, and if the dark baryon asymmetry is around $Y_\Bd \simeq 10^{-9}$. 
This result illustrates the same ``coincidence'' that comes up in models of asymmetric dark matter~\cite{Nussinov:1985xr,Kaplan:2009ag} where the dark matter's mass and asymmetry are comparable to the baryon's mass and asymmetry.  

%=============
\begin{figure}[t]
\begin{center}
\includegraphics[width=0.48\textwidth]{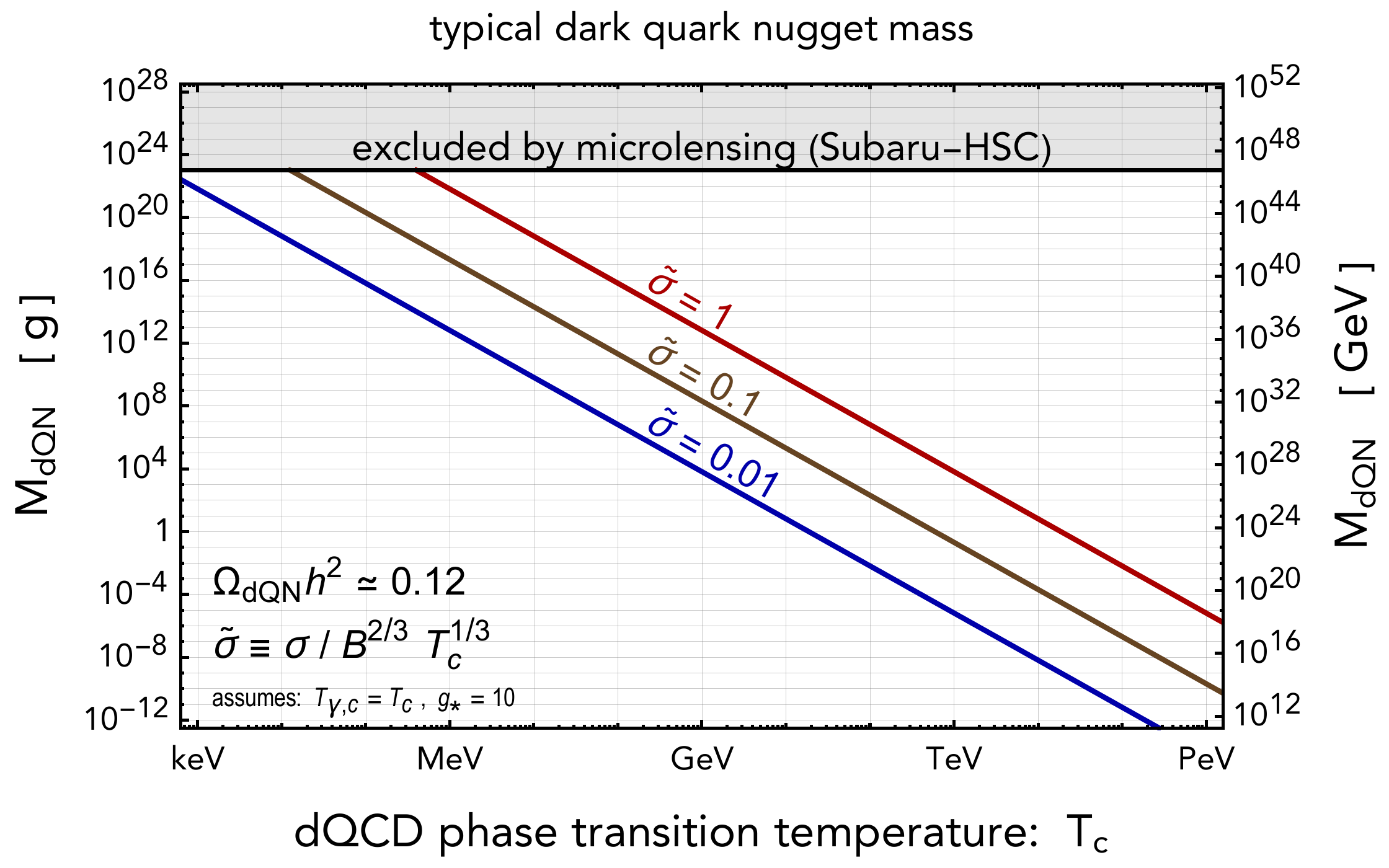} \hfill
\includegraphics[width=0.49\textwidth]{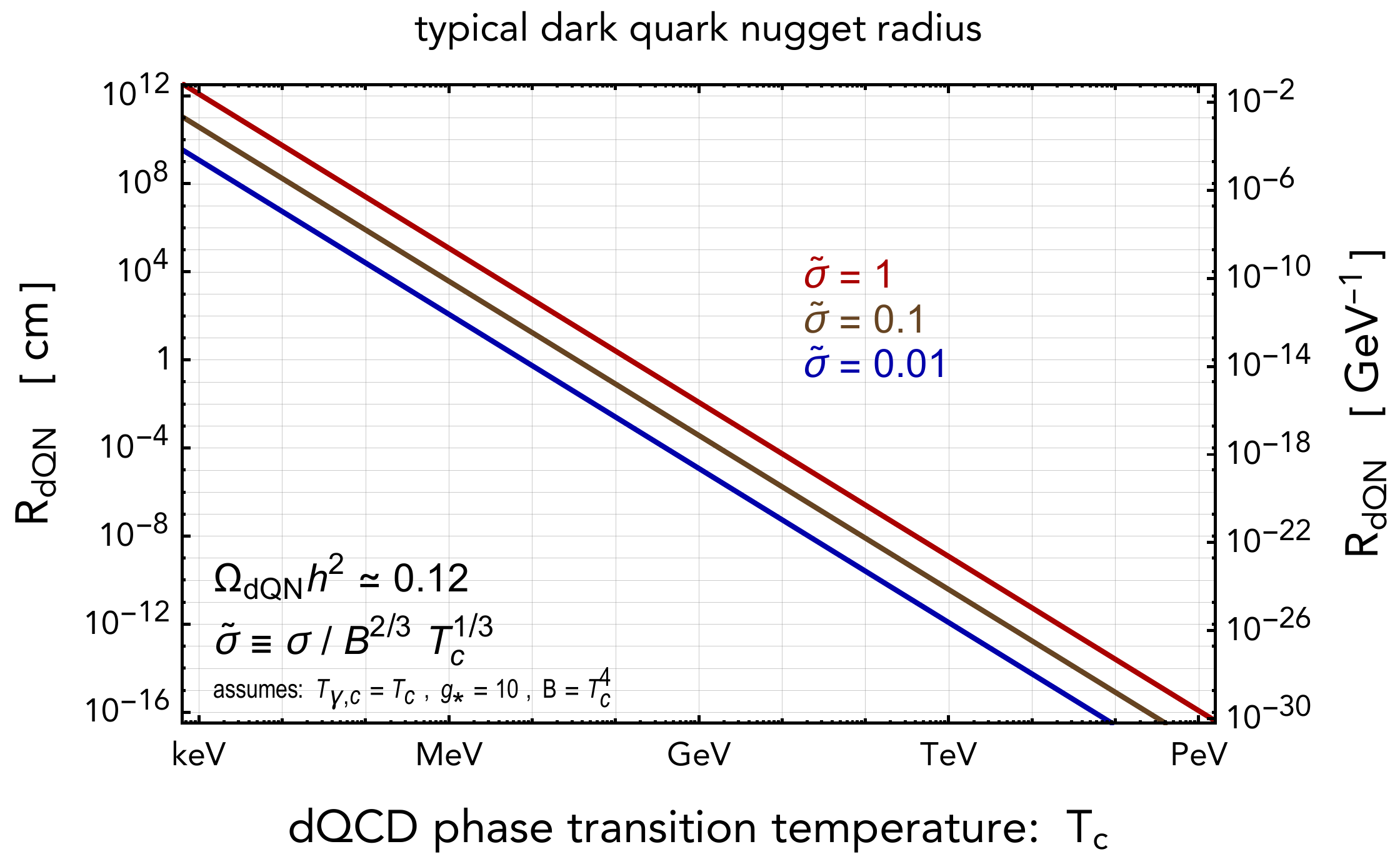} 
\caption{\label{fig:MdQN}
The typical mass (left panel) and radius (right panel) of a dark quark nugget are shown here as functions the critical temperature of the confining phase transition.  We assume $T_\gamma(t_c) = T_d(t_c) \equiv T_c$, but if the dark sector is colder then the mass and radius are reduced according to \erefs{eq:R_dQN_alt}{eq:M_dQN_alt}.  The dimensionless parameter $\tilde{\sigma} \equiv \sigma / (B^{2/3} T_c^{1/3})$ measures the surface tension of the confined-phase bubbles at the time of formation, which affects the initial dQN density through \eref{eq:n_dQN_tc}.  The dark quark nuggets are assumed to occupy the majority of dark matter energy density.  If the scale of the confining phase transition is larger than $\sim 10 \TeV$ then free dark baryons over close the universe; see the discussion in \sref{sec:dark_baryons}. Also shown is the Subaru-HSC microlensing constraint after taking the wave effects into account~\cite{Niikura:2017zjd,Katz:2018zrn}. 
}
\end{center}
\end{figure}

%=============
Solving \eref{eq:Omega_dQN} for $Y_\Bd$ lets us write \erefs{eq:R_dQN}{eq:M_dQN} as 
\begin{align}
	R_\dQN & \simeq
	\bigl( 0.081~\mathrm{cm} \bigr) 
	\left( \frac{\Omega_\dQN h^2}{0.12} \right)^{1/3} 
	\left[ \frac{B}{(0.1 \GeV)^4} \right]^{-1/3} 
	\left( \frac{\Tc}{0.1 \GeV} \right)^{-1} 
	\left( \frac{\tilde{\sigma}}{0.1} \right)^{3/2} \label{eq:R_dQN_alt}
	\com \\ 
	M_\dQN & \simeq 
	\bigl( 2.1 \times 10^{11} \gram \bigr) 
	\left( \frac{\Omega_\dQN h^2}{0.12} \right) 
	\left( \frac{\Tc}{0.1 \GeV} \right)^{-3} 
	\left( \frac{\tilde{\sigma}}{0.1} \right)^{9/2} \label{eq:M_dQN_alt}
	\per
\end{align} 
In \fref{fig:MdQN} we show the dark quark nugget's mass and radius for the interesting range of phase transition temperatures from $T_{\gamma,c} = 1 \keV$ to $1 \PeV$.

%==================================
% Signatures and testable predictions
%==================================
\section{Signatures and testable predictions}\label{sec:signatures}

%=========
In this section we discuss various observational signatures of the theory that we have presented above.  
Some of these observables directly test for the presence of dark quark nuggets in our universe while other indirectly probe the dark QCD model.

%--------------------------------------------
% Dark radiation
%--------------------------------------------
\subsection{Dark radiation}\label{sec:dark_radiation}

%=========
In addition to a dark matter candidate, the dark QCD model also admits a dark radiation candidate. The presence of dark radiation in the universe is felt through its gravitational influence, particularly during the formation of the cosmic microwave background (CMB).  
In this section we discuss how CMB observations lead to constraints on the dark QCD model and its dark radiation.  

%=========
In general we can write the energy density of particles in the dark sector as 
\begin{align}
	\rho_d = \rho_{d,\mathrm{rad}} + \rho_{d,\mathrm{mat}}  \,,
\end{align}
where $\rho_{d,\mathrm{rad}}$ is the energy density of (relativistic) dark radiation and $\rho_{d,\mathrm{mat}}$ is the energy density of (nonrelativistic) dark matter.  
The various particle species in the dark sector -- quark and gluons in the unconfined phase and mesons and baryons in the confined phase -- are distributed between radiation and matter.  

%=========
In the following discussion we consider the model with $m_i = 0$ in \eref{eq:Lagrangian}, which corresponds to massless dark quarks in the unconfined phase and massless dark mesons (Goldstone bosons) in the confined phase.\footnote{If these masses were nonzero, it may be possible to evade the constraints on dark radiation by allowing the dark mesons to decay to visible-sector particles.  However, relaxing the assumption $m_i = 0$ opens an additional layer of model building that we do not seek to address at this time.}  
If all species of particles in the dark sector are in thermal equilibrium at a common temperature $T_d$ then the energy densities in the dark sector are given by\footnote{The factor $2(N_d^2-1)$ counts the two spin states of the $(N_d^2-1)$ species of dark gluons; the factor $4N_dN_f$ counts the two spin states of the $N_d N_f$ species of dark quarks and antiquarks; and the factor $(N_f^2-1)$ or $(2N_f^2 - N_f - 1)$ counts the flavors of massless dark mesons. }
\begin{align}
	\rho_{d,\mathrm{rad}} & = 
	\frac{\pi^2}{30} \, g_{\ast,d} \, T_d^4 
	\ \ , \ \ \ \
	g_{\ast,d}
	= \begin{cases}
	2 (N_d^2 - 1) + \frac{7}{8} (4 N_d N_f)  \quad \qquad , \quad \text{unconfined phase} \\
	\begin{cases} 
	(N_f^2 - 1)~\mbox{for}~N_d \geq 3 & , \quad \text{confined phase} \,, \\ 
	(2N_f^2 - N_f - 1)~\mbox{for}~N_d =2 & , \quad \text{confined phase} \\
	\end{cases} 
	\end{cases} \label{eq:rho_d_rad} \\ 
	\rho_{d,\mathrm{mat}} & = \begin{cases}
	0 & , \quad \text{unconfined phase} \\ 
	\rho_\Bd + \rho_{\overline{\mathsf{B}}_{\rm d}} + \rho_\dQN & , \quad \text{confined phase} \\ 
	\end{cases} \label{eq:rho_d_mat} 
	\per
\end{align} 
The first equality also defines the effective number of relativistic species in the dark sector, denoted by $g_{\ast,d}$.  
The terms in $\rho_{d,\mathrm{mat}}$ count the energy density of non-relativistic species carrying dark baryon number, which includes dark baryons, dark antibaryons, and dark quark nuggets.

%=========
When placing constraints on dark radiation, it is customary to compare the dark radiation energy density against the energy density of a single, massless neutrino/antineutrino pair, $\rho_{\nu1} = (2)(7/8) (\pi^2/30)  T_\nu^4$ where $T_\nu = (4/11)^{1/3} \, T_\gamma$ at the CMB epoch~\cite{Kolb:1990}. 
Thus the dark radiation is parametrized by $\Delta N_\mathrm{eff} \equiv \rho_{d,\mathrm{rad}} / \rho_{\nu1} |_{t_\mathrm{cmb}}$, which evaluates to 
\begin{align}\label{eq:DNeff}
	\Delta N_\mathrm{eff} 
	= \left( \frac{11}{4} \right)^{4/3} \left( \frac{4}{7} \right) g_{\ast,d}(t_\mathrm{cmb}) \, \frac{T_d(t_\mathrm{cmb})^4}{T_\gamma(t_\mathrm{cmb})^4} 
	\per
\end{align}
In general the dark and visible sectors may have different temperatures.  
The parameter $\Delta N_\mathrm{eff}$ is already strongly constrained~\cite{Aghanim:2018eyx}, due to the absence of evidence for dark radiation at the CMB epoch, and next-generation observations~\cite{Abazajian:2016yjj} are projected to improve the sensitivity by an order of magnitude: 
\begin{align}\label{eq:DNeff_constraints}
	\Delta N_\mathrm{eff} & < 0.2 \quad \text{at 95\% C.L.}  \,,  && \text{current limit -- Planck~2018}  \,, \\ 
	\sigma(\Delta N_\mathrm{eff}) & = 0.03 \,, && \text{projected sensitivity -- CMB-S4} \nonumber
	\per
\end{align}
The presence of dark radiation at the epoch of nucleosynthesis is more weakly constrained, $\Delta N_\mathrm{eff} < 1$ at 95\% C.L.~\cite{Mangano:2011ar}.

%=========
To make a prediction for $\Delta N_\mathrm{eff}$ we must estimate $T_d / T_\gamma$, but this ratio depends on the history of interactions between the dark and visible sectors. Without loss of generality, we identify three scenarios.  

%=========
\paragraph{1.  The dark and visible sectors are thermalized at the CMB epoch.} 
If the dark sector remains in thermal equilibrium with the visible sector at the CMB epoch, then we take $T_d = T_\gamma$ in \eref{eq:DNeff} to evaluate $\Delta N_\mathrm{eff}$.  
We can distinguish two cases, either:  1a) the dark sector is still in the unconfined phase at $t_\mathrm{cmb}$ or 1b) it is in the confined phase.  
For case (1a) we find $\Delta N_\mathrm{eff} \gg 1$ for any $N_d \geq 2$ and $N_f \geq 1$.  
For case (1b) we have $\Delta N_\mathrm{eff} \gg 1$ for any $N_d \geq 2$ and $N_f \geq 2$, but $\Delta N_\mathrm{eff} = 0$ if $N_f = 1$, because there is no Goldstone boson.  
Nevertheless, a model with $N_f = 1$ is not expected to have a first-order phase transition~\cite{Pisarski:1983ms} or allow for the formation of dQNs.  
In light of the constraints on $\Delta N_\mathrm{eff}$ in \eref{eq:DNeff_constraints}, this first scenario is not viable.  

%=========
\paragraph{2. The dark and visible sectors decouple prior to the CMB epoch.}  
The $\Delta N_\mathrm{eff}$ constraints are relaxed if the dark sector decoupled from the Standard Model at a time $t_\mathrm{dec} < t_\mathrm{cmb}$, before the CMB epoch.  
If we assume that the cosmological expansion causes the two sectors to cool adiabatically,\footnote{The adiabatic cooling assumption breaks down if the dark QCD phase transition occurs abruptly, because the liberated latent heat will heat the dark plasma.  We neglect this effect for these estimates.  } then the comoving entropy density is separately conserved in the two sectors, and we can write 
\begin{subequations}\label{eq:adiabat}
\begin{align}
	a(t)^3 \, g_{\ast,d}(t) \, T_d(t)^3 & = a(t_\mathrm{dec})^3 \, g_{\ast,d}(t_\mathrm{dec}) \, T_d(t_\mathrm{dec})^3  \,, \\
	a(t)^3 \, g_{\ast,\gamma}(t) \, T_\gamma(t)^3 & = a(t_\mathrm{dec})^3 \, g_{\ast,\gamma}(t_\mathrm{dec}) \, T_\gamma(t_\mathrm{dec})^3  \,.
\end{align}
\end{subequations}
Here $g_{\ast,d}(t)$ denotes the effective number of relativistic species in the dark sector at time $t$, and it is given by \eref{eq:rho_d_rad}.  
Similarly $g_{\ast,\gamma}(t)$ denotes the effective number of relativistic species in the visible sector (Standard Model degrees of freedom).  
Assuming no new light degrees of freedom beyond the Standard Model and the dark QCD, then this factor is as large as $g_{\ast,\gamma} = 106.75$ for $T_\gamma \gtrsim 160 \GeV$ before electroweak symmetry breaking, and it decreases to $g_{\ast,\gamma} = 3.91$ for $T_\gamma \lesssim 0.2 \MeV$ after neutrino scattering and electron-positron annihilations have frozen out.  
At the time of decoupling $T_d(t_\mathrm{dec}) = T_\gamma(t_\mathrm{dec})$, but as particle species go out of equilibrium the temperatures will begin to differ.  
Solving \eref{eq:adiabat} for $t_\mathrm{dec} < t$ gives 
\begin{align}\label{eq:T_d}
	\frac{T_d(t)}{T_\gamma(t)} = \left[ \frac{g_{\ast, \gamma}(t)}{g_{\ast, \gamma}(t_\mathrm{dec})} \right]^{1/3} \, \left[ \frac{g_{\ast, d}(t)}{g_{\ast, d}(t_\mathrm{dec})} \right]^{-1/3} 
	\com
\end{align}
and \eref{eq:DNeff} becomes 
\begin{align}\label{eq:DNeff_final}
	\Delta N_\mathrm{eff} 
	& \simeq \bigl( 0.027 \bigr) \bigl[ g_{\ast,d}(t_\mathrm{cmb}) \bigr]^{-1/3} \bigl[ g_{\ast,d}(t_\mathrm{dec}) \bigr]^{4/3} \left[ \frac{g_{\ast,\gamma}(t_\mathrm{cmb})}{3.91} \right]^{4/3} \left[ \frac{g_{\ast,\gamma}(t_\mathrm{dec})}{106.75} \right]^{-4/3} 
	\per
\end{align}
Formulas for $g_{\ast, d}$ appear in \eref{eq:rho_d_rad}.  

%=============
\begin{figure}[h!]
\begin{center}
\includegraphics[width=0.95\textwidth]{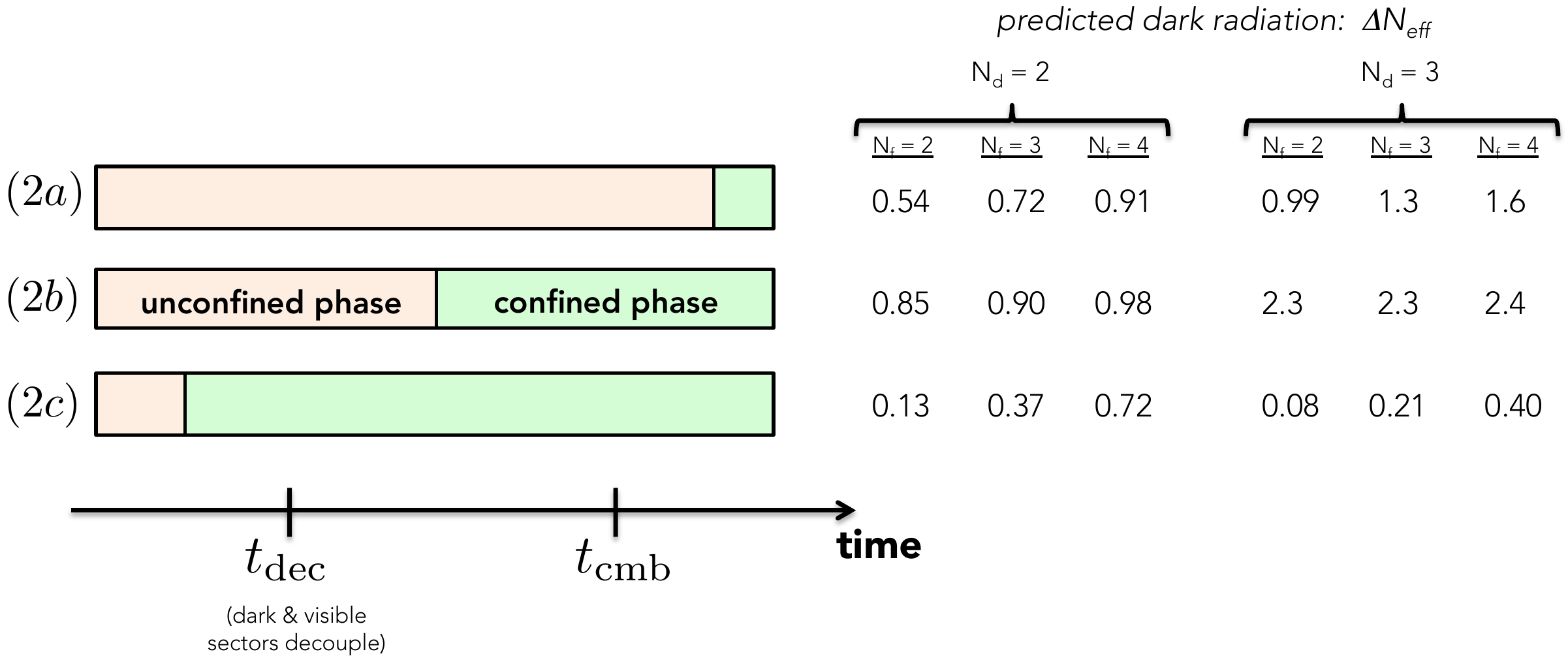} 
\caption{\label{fig:dark_rad}
The predicted dark radiation, parametrized by $\Delta N_\mathrm{eff}$, is shown for the three cases depending on whether the dark sector is in the unconfined or the confined phase at the time when it thermally decouples from the visible sector ($t_\mathrm{dec}$) and the time when the CMB is generated ($t_\mathrm{cmb}$).  Observational constraints \pref{eq:DNeff_constraints} strongly prefer case (2c) in which the confining phase transition occurs while the dark and visible sectors are still in thermal equilibrium. We assume that decoupling occurs before the electroweak epoch with $g_{\ast,\gamma} = 106.75$, and otherwise $\Delta N_\mathrm{eff}$ is larger according to \eref{eq:DNeff_final}.  We also assume massless dark mesons, but if the dark mesons are instead allowed to decay to SM particles before $t_\mathrm{cmb}$ then the predicted $\Delta N_\mathrm{eff}$ is smaller.  For $N_f = 1$ there is no dark radiation for cases (2b) and (2c).
}
\end{center}
\end{figure}

%=========
One can now distinguish three different cases: 2a) the dark sector is thermally decoupled while in the unconfined phase and it remains in the unconfined phase at the CMB epoch, 2b) the dark sector is thermally decoupled while in the unconfined phase and it passed into the confined phase prior to the CMB epoch, and 2c) the dark sector is thermally decoupled while in the confined phase and it remains in the confined phase at the CMB epoch.  
These cases are illustrated in \fref{fig:dark_rad}.  
For each of these three cases, the predicted $\Delta N_\mathrm{eff}$ is given by 
\begin{align}\label{eq:DNeff_cases}
	\Delta N_\mathrm{eff} = \begin{cases}
	0.027 \, \bigl[ 2 (N_d^2 - 1) + \frac{7}{8} (4 N_d N_f) \bigr] & , \quad \text{(2a)} \\ 
	\begin{cases} 
	0.027 \, \frac{[ 2 (N_d^2 - 1) + \frac{7}{8} (4 N_d N_f) ]^{4/3}}{[ N_f^2 - 1 ]^{1/3} } & , \quad N_d \geq 3 \\ 
	0.027 \, \frac{[ 2 (N_d^2 - 1) + \frac{7}{8} (4 N_d N_f) ]^{4/3}}{[ 2N_f^2 - N_f - 1 ]^{1/3} } & , \quad N_d = 2 
	\end{cases} & , \quad \text{(2b)} \\ 
	\begin{cases} 
	0.027 \, \bigl[ N_f^2 - 1 \bigr] & , \quad N_d \geq 3 \\ 
	0.027 \, \bigl[ 2N_f^2 - N_f - 1 \bigr] & , \quad N_d = 2 
	\end{cases} & , \quad \text{(2c)} \\ 
	\end{cases} 
	\per
\end{align}
Here we have chosen $g_{\ast,\gamma}(t_\mathrm{dec}) = 106.75$, but if decoupling occurs after the electroweak epoch ($T_\mathrm{ew}$) instead, then the value of $g_{\ast,\gamma}(t_\mathrm{dec})$ is smaller and $\Delta N_\mathrm{eff}$ is even larger, as can be seen from \eref{eq:DNeff_final}.
For cases (2a) and (2b), the predicted $\Delta N_\mathrm{eff}$ is always larger than the level of the observational constraints \pref{eq:DNeff_constraints}, mostly due to the large number of gluon degrees of freedom, {\it i.e.} the $2(N_d^2-1)$ term with $N_d \geq 3$.  
However for case (2c), in which the dark sector is already confined when it decouples from the visible sector, we predict an acceptable level of dark radiation for the model with $N_d = N_f = 2$ and for the models with $N_d \geq 3$ and $N_f = 2$ or $3$.  
Since we also need $N_f \geq 3$ to ensure a first order phase transition (see the discussion in \sref{sec:dark_quark_matter}), the only viable models are
\begin{align}
	N_d \geq 3, \qquad N_f = 3, \qquad T_\mathrm{ew} < T_\mathrm{dec} < T_c, \qquad \Delta N_\mathrm{eff} \simeq 0.21 \,,
\end{align}
in order to generate quark nuggets while avoiding constraints from dark radiation.  
Alternatively, it may be possible to open up the parameter space by lifting the dark meson mass and allowing it to decay to Standard Model particles before the CMB epoch.  

%=============
\begin{figure}[ht!]
\begin{center}
\includegraphics[width=0.65\textwidth]{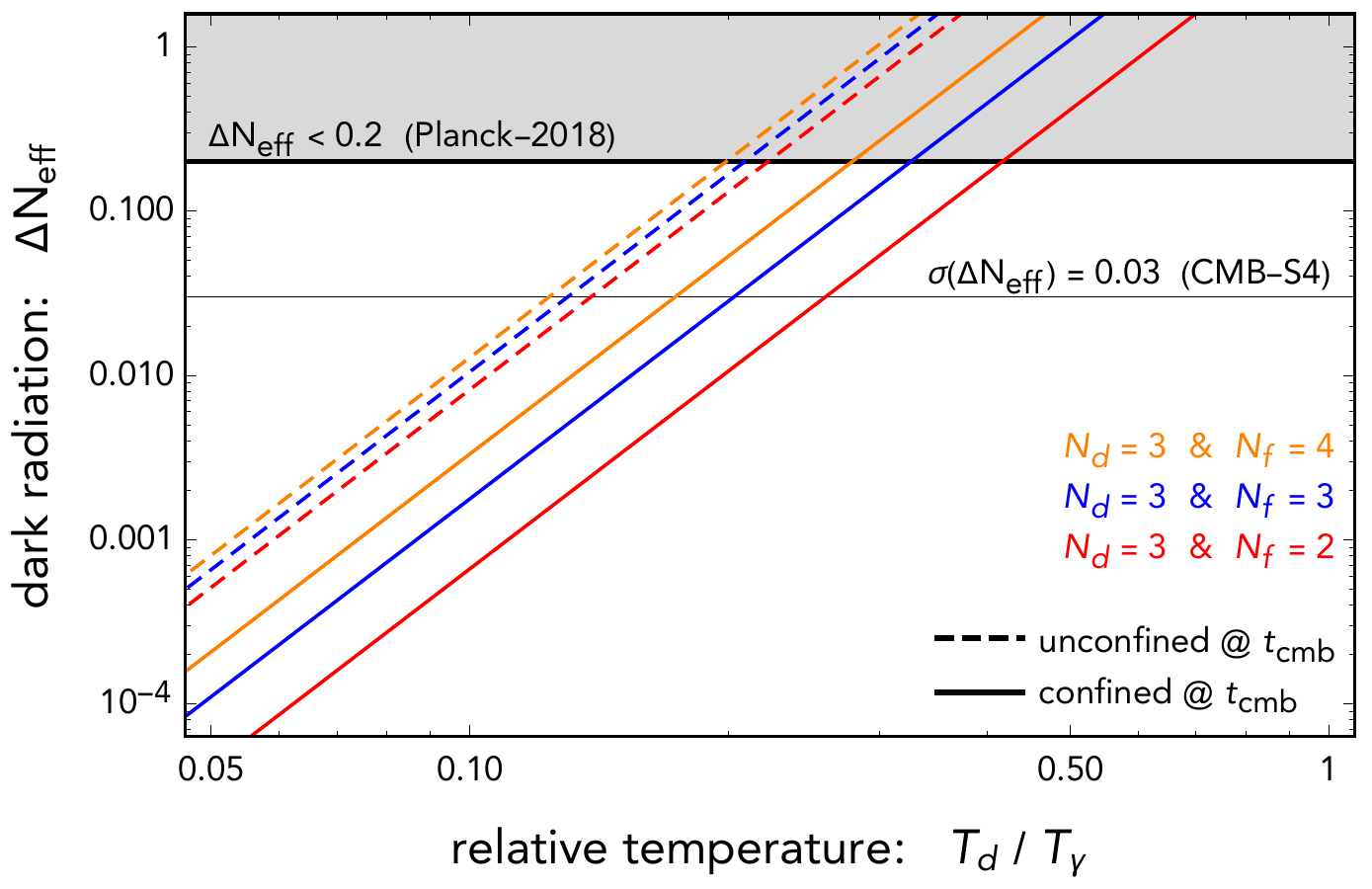} 
\caption{\label{fig:temp_ratio}
Here we show the predicted dark radiation, parametrized by $\Delta N_\mathrm{eff}$, for case (3) in which the dark sector is never thermalized with the SM and the temperature ratio, $T_d / T_\gamma$, is determined by initial conditions.  Several values of $N_d$ and $N_f$ are shown, and we consider two cases depending on whether or not the dark sector is confined at the CMB epoch.  Provided that $T_d \lesssim T_\gamma / 3$ the dark radiation is small enough to evade existing limits, and if $T_d \gtrsim T_\gamma / 10$ then the next-generation CMB-S4 program may uncover evidence for dark radiation. 
}
\end{center}
\end{figure}

%=========
\paragraph{3.  The dark and visible sectors never thermalize.} 
If the dark sector never reaches thermal equilibrium with the Standard Model, and if the freeze-in population is negligible (see also \rref{Adshead:2016xxj}), then the ratio $T_d/T_\gamma$ is controlled by the physics that populated the dark and visible sectors initially.  For instance if both sectors are populated directly from decay of the inflaton field $\phi$ after cosmological inflation has evacuated the observable universe~\cite{Albrecht:1982mp,Dolgov:1982th,Abbott:1982hn}, then $T_d / T_\gamma$ is proportional to a ratio of branching fractions $\mathrm{BF}(\phi \to \mathrm{dark}) \, / \, \mathrm{BF}(\phi \to \mathrm{SM})$.  The ratio $T_d / T_\gamma$ can be made arbitrarily small in a model in which the inflaton decays predominantly to the visible sector, and the constraints from $\Delta N_\mathrm{eff}$ can be avoided.  In \fref{fig:temp_ratio}, we show the predicted dark radiation as a function of the temperature ratio $T_d/T_\gamma$.  Even a small splitting, $T_d / T_\gamma \sim 1/3$, is enough to evade existing constraints, but still provide a target for next-generation surveys.  However, if the two sectors do not thermalize, then the dark and visible baryon asymmetries may either arise directly from the inflaton decay (if it is $\mathcal{CP}$- and baryon-number violating), or baryogenesis may occur separately in the two sectors.

%--------------------------------------------
% Free dark baryons and antibaryons
%--------------------------------------------
\subsection{Free dark baryons and antibaryons}\label{sec:dark_baryons}

%=========
After the confining phase transition occurs, the dark baryon number is carried by the dark baryons ($B_d$), the dark antibaryons ($\bar{B}_d$), and the dark quark nuggets ($\dQN$)
\footnote{The free dark baryons may undergo dark nucleosynthesis to form dark nuclei, and this idea has been explored recently by several authors~\cite{Hardy:2014mqa,Wise:2014jva,Wise:2014ola,Krnjaic:2014xza,Gresham:2017cvl}. Since the dark baryons typically make up a sub-dominant population of the dark matter, the total dark matter relic abundance is approximately not affected. Also, the dark baryon number for the dark nucleus coagulation is dramatical smaller than the one in nuggets, and their detection potential could be dramatically different from nuggets. 
}
%There may be interesting implications for direct detection prospects, but we leave a detailed exploration of the phenomenology for future work.}}.  
In this section we estimate the relic abundances of the dark baryons and antibaryons.  
We assume that dark baryon number is conserved, which forbids the dark baryons/antibaryons from decaying, and instead they contribute to the dark matter.  

%=========
The dark baryons and antibaryons are kept in thermal equilibrium with the dark mesons, such as the dark pions $\pi_d$, through annihilation reactions such as $B_d + \bar{B}_d \leftrightarrow \pi_d + \pi_d$ and multi-meson final states.  
Let $\langle \sigma v \rangle$ denote the thermally-averaged cross section for this annihilation reaction.  
At temperatures below the mass of the dark baryon/antibaryon, $T \ll m_\dB$, the thermally averaged cross section is well approximated by 
\begin{align}\label{eq:sigma_annihilation}
	\langle \sigma v \rangle \approx (50\,\mbox{mb}\cdot\mbox{c})\, \left( \frac{1 \GeV}{m_\dB}\right)^2
	\com
\end{align}
where we have used the low-$\beta$ $\bar{p}p$ annihilation rates~\cite{Zenoni:1999st}.  
This is roughly $\langle \sigma v \rangle \simeq 130 / m_\dB^2$.  

%=========
If the dark baryon asymmetry is negligibly small then the relic abundances of dark baryons and antibaryons, $\Omega_{B_d}$ and $\Omega_{\bar{B}_d}$, are controlled by thermal freeze out, which occurs when the plasma temperature in the dark sector is approximately $T_d(t_\mathrm{fo}) \simeq m_\dB / 20$.  
The standard freeze out calculation~\cite{Kolb:1990} gives the relic abundances to be 
\begin{align}\label{eq:Omega_Bd_1}
	\Omega_{B_d} h^2 = \Omega_{\bar{B}_d} h^2 
	& \simeq \bigl( 0.052 \bigr) \left( \frac{\langle \sigma v \rangle}{130 \, m_\dB^{-2}} \right)^{-1} \left( \frac{m_\dB}{200 \TeV} \right)^{2} \left( \frac{m_\dB / T_d(t_\mathrm{fo})}{20} \right) \left( \frac{T_d(t_\mathrm{fo})}{T_\gamma(t_\mathrm{fo})} \right) \left( \frac{g_\ast}{100} \right)^{-1/2}
	\per
\end{align}
The factor of $T_d(t_\mathrm{fo}) / T_\gamma(\mathrm{fo}) \leq 1$ arises because the dark and visible sectors may be thermally decoupled at the time of dark baryon freeze out.
However, as we have already discussed in \sref{sec:nuggets}, a dark-baryon-number asymmetry is required for the formation of dark quark nuggets, and this asymmetry may affect the relic abundance of free dark baryons and antibaryons as well (as we encounter in models of asymmetric dark matter~\cite{Nussinov:1985xr,Kaplan:2009ag}).  
Recall from \eref{eq:f_nug} that the fraction of dark baryon number carried by the free dark baryons is $f_\mathrm{free} Y_\Bd$ where $f_\mathrm{free} = 1 - f_\mathrm{nug} \ll 1$ is desirable for the formation of nuggets.  
If the dark baryon asymmetry is large enough, then the relic abundances are given by
\begin{align}\label{eq:Omega_Bd_2}
	\Omega_{B_d} h^2 
	\simeq \bigl( 0.14 \bigr) \left( \frac{m_\dB}{50 \GeV} \right) \left( \frac{1-f_\mathrm{nug}}{0.01} \right) \left( \frac{Y_\Bd}{10^{-9}} \right) 
	\qquad \text{and} \qquad 
	\Omega_{\bar{B}_d} h^2 \approx 0 
	\com
\end{align}
which is insensitive to $\langle \sigma v \rangle$.  
If $Y_\Bd < 0$ then the expressions for $\Omega_{B_d}$ and $\Omega_{\bar{B}_d}$ are exchanged.  
For sure, since dark quark nuggets have the energy density with a factor of around $f_\mathrm{nug}/(1-f_\mathrm{nug})$ larger than that from free dark baryons, the specific parameter choice of $m_\dB=50$~GeV and $Y_\Bd=10^{-9}$ will have dark matter overclose the universe. 

\begin{figure}[t]
\begin{center}
\includegraphics[width=0.68\textwidth]{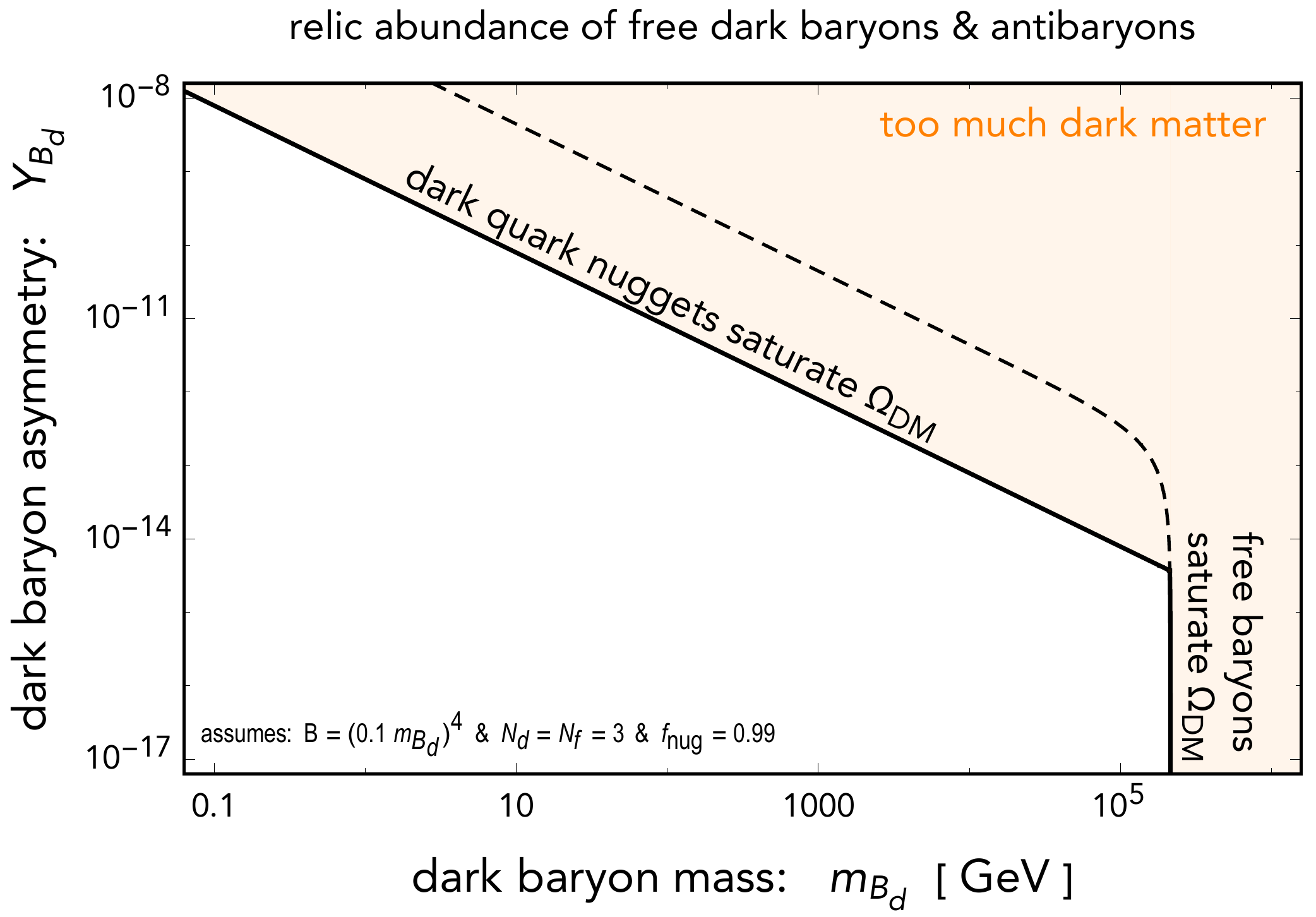} \hfill
\caption{\label{fig:dark_baryon}
The relic abundances of free dark baryons and antibaryons are shown here in comparison with the relic abundance of dark quark nuggets.  Note that the curve for free dark baryons \pref{eq:Omega_Bd_2} scales as $f_\mathrm{free} = 1 - f_\mathrm{nug}$, whereas the curve for dark quark nuggets scales as $f_\mathrm{nug}$; we have taken $f_\mathrm{nug} = 0.99$ for illustration, but this value may vary greatly across models. For the free dark baryon thermal relic abundance, we have used $T_d(t_\mathrm{fo})=T_\gamma(t_\mathrm{fo})$. 
}
\end{center}
\end{figure}

%=========
The relic abundance of free dark baryons is shown in \fref{fig:dark_baryon} as a function of the dark baryon mass scale and the dark baryon asymmetry.  
Requiring the relic abundance of dark baryons to be smaller than the observed density of dark matter, $\Omega_\DM h^2 \simeq 0.12$, yields an upper bound~\cite{Griest:1989wd} of $m_\dB \lesssim 200 \TeV$.  
Recall from \eref{eq:mass_condit} that we need $T_c \lesssim m_\dB / 7$ to ensure that nuggets are able to form, and therefore the over-closure condition implies an upper bound on the dark-sector temperature at the phase transition:  
\begin{align}\label{eq:Tc_upper}
	\Omega_{B_d} + \Omega_{\bar{B}_d} < \Omega_\DM 
	\qquad \Rightarrow \qquad 
	T_c \lesssim 30 \TeV
	\per
\end{align}
However, the temperature in the dark sector may be smaller than the temperature in the visible sector, $T_c \leq \Tc$, which affects the corresponding lower bounds on the dQN mass and radius through \erefs{eq:R_dQN_alt}{eq:M_dQN_alt}.  

%=========
For comparison \fref{fig:dark_baryon} also shows the relic abundance of dark quark nuggets \pref{eq:Omega_dQN}.  
For $m_\dB \lesssim 200 \TeV$ the relative abundances are given by 
\begin{align}
	\frac{\text{free dark baryons}}{\text{dark quark nuggets}}: \qquad \frac{\Omega_{B_d} + \Omega_{\bar{B}_d}}{\Omega_\dQN} \simeq \bigl( 0.031 \bigr) \left( \frac{N_f^{1/4}}{N_d^{3/4}} \right) \, \left( \frac{m_\dB/B^{1/4}}{10} \right) \, \left(\frac{f_\mathrm{free}/f_\mathrm{nug}}{0.01}\right) 
	\per
\end{align}
Note that the free dark baryons are a subdominant population of the dark matter provided that 
\begin{align}\label{eq:f_nug_limit}
	f_\mathrm{nug} = 1 - f_\mathrm{free} > \left[ 1 + 3.3 \, \frac{N_d^{3/4} B^{1/4}}{N_f^{1/4} m_\dB} \right]^{-1}
	\com
\end{align}
which evaluates to $f_\mathrm{nug} > 0.636$ for $N_d = N_f = 3$ and $m_\dB = 10 B^{1/4}$.  
An expression for $f_\mathrm{nug}$ appears in \eref{eq:f_nug}, and by comparing with the limit above, we find that free dark baryons typically make up a subdominant component of the dark matter, which is predominantly composed of dark quark nuggets.

%=========

%=========
Since the free dark baryons and antibaryons are very abundant, it may be possible to detect their presence with direct detection experiments on Earth.  
Their gravitational influence is expected to be exceedingly weak, and therefore an additional, direct coupling between the dark sector and the SM is required.  
The nature of this interaction depends on (as yet unspecified) UV physics.  
As an example we will use the vector-vector interactions, $\overline{\psi}_{d,L} \gamma_\mu \psi_{d,L} \, \overline{d}_R \gamma^\mu d_R /\Lambda_\mathrm{UV}^2$, which could be generated by integrating out a heavy scalar coupling to both a dark quark and an ordinary quark and using the Fierz transformation.  
Then the matrix element for spin-independent (SI) scattering of a dark baryon off a proton or neutron is written as $\mathcal{M}_{p,n} = J_{\psi_d}^0 J_{p, n}^0/(4 \Lambda^2_\mathrm{UV})$ where $J_{\psi_d}^0= \langle B_d |  \overline{\psi}_d\gamma^0 \psi_d | B_d \rangle \approx N_d$ and $J_{p, n}^0 = \langle p, n | \overline{d}\gamma^0 d | p, n\rangle \approx 1, 2$.  
For a Fermionic dark baryon, the SI scattering cross section for a neutron is
\beqa\label{eq:sigma_SI}
\sigma^\mathrm{SI}_{B_d - n} = \frac{N_d^2\,\mu^2_{B_d - n}}{4\pi\,\Lambda^4_\mathrm{UV}} \simeq \bigl( 2.5 \times 10^{-44} \cm^2 \bigr) \left( \frac{\Lambda_\mathrm{UV}}{10 \TeV} \right)^{-4} \left( \frac{N_d}{3} \right)^2 
\com
\eeqa
where $\mu_{B_d - n} = m_\dB m_n / (m_\dB + m_n) \approx m_n$ is the reduced mass for $m_\dB \gg m_n$.  
Recent null results from the one tonne-year exposure of XENON1T~\cite{Aprile:2018dbl}, implies an upper bound on the dark baryon scattering cross section at the level of $\sigma_{B_d-n}^\mathrm{SI} \lesssim (4.1\times 10^{-47} \cm^2) (m_\dB/30\GeV) [\Omega_\dQN / (\Omega_{B_d} + \Omega_{\bar{B}_d})]$, where the $\Omega$-factor arises because dark baryons are only a subdominant component of the dark matter.  
Thus the non-observation of free dark baryons by XENON1T imposes 
\begin{align}
	\Lambda_\mathrm{UV} 
	\gtrsim \bigl( 42 \TeV \bigr) 
	\left( \frac{N_d}{3} \right)^{5/16} 
	\left( \frac{N_f}{3} \right)^{1/16} 
	\left( \frac{B}{(0.1 \GeV)^4} \right)^{-1/16} 
	\left( \frac{f_\mathrm{free} / f_\mathrm{nug}}{0.01} \right)^{1/4} 
	\per
\end{align}
This limit also means that if the cutoff scale is not too far from $40 \TeV$, the future results from direct detection experiments could have a chance to discovery the dark baryon.

%--------------------------------------------
% Stochastic gravitational wave background
%--------------------------------------------
\subsection{Stochastic gravitational wave background}\label{sec:grav_wave}

%=========
It is well known that cosmological phase transitions can generate a stochastic background of gravitational waves (GW) if the transition is first order~\cite{Kamionkowski:1993fg}.  
First order phase transitions in dark sectors have also been studied specifically; see e.g.~Refs.~\cite{Schwaller:2015tja,Jaeckel:2016jlh,Tsumura:2017knk,Addazi:2017gpt,Aoki:2017aws,Huang:2017laj,Baldes:2018emh,Croon:2018erz}.  
In general, three processes contribute to the stochastic GW background during a first-order phase transition: the collision of the scalar field bubbles, sound waves in the plasma, and the magnetohydrodynamic (MHD) turbulence.  
The total GW spectrum is then well approximated by the linear sum of these three contributions:
\begin{align}
	\Omega_\mathrm{gw}h^2 \approx \Omega_\phi h^2 + \Omega_\mathrm{sw}h^2 + \Omega_\mathrm{turb}h^2 
	\per
\end{align}
The spectra of these three sources are determined by several key parameters from the bubble nucleation process.  
The parameter $\beta^{-1}$ measures the duration of the phase transition, and it is customary to write the dimensionless ratio $\beta/H$ where $H$ is the Hubble parameter at the time when GWs are generated; see also \eref{eq:beta_def}.  
We assume that the universe is radiation dominated during the phase transition with the dominant energy component having a temperature $T_\ast \approx \Tc$.  
The dimensionless parameter $\alpha$ measures the released vacuum energy as compared to the radiation energy of the plasma after the phase transition is completed; see also \eref{eq:alpha_def}.  
The parameter $\alpha$ also controls the efficiency with which energy is transferred into the bulk motion of the fluid; this efficiency is parametrized by $\kappa_\mathrm{f}$, and an explicit expression appears below.  
The parameter $v_w$ measures the speed of the bubble wall in the rest frame of the plasma.  

%=========
For bubbles that reach a terminal velocity (rather than ``running away''), the contribution to gravitational waves from the bubble collisions themselves has been shown by recent numeric study to be negligible~\cite{Hindmarsh:2015qta}.
The GW signal from MHD turbulence also turns out to be negligible for the parameter range we are considering.  
Therefore we only present the formula for the sound wave contribution, which fits to~\cite{Hindmarsh:2015qta}
\begin{align}\label{eq:sw_spectrum}
\Omega_\mathrm{sw}h^2 = \bigl(8.5 \times 10^{-6} \bigr) \left(\frac{g_\ast}{100}\right)^{-1/3}\Gamma^2 \,\overline{U}_\mathrm{f}^4 \,\left(\frac{\beta}{H}\right)^{-1} v_\mathrm{w} \left(\frac{f}{f_\mathrm{sw}}\right)^3 \left(\frac{7}{4+3(f/f_\mathrm{sw})^2} \right)^{7/2}
	\per
\end{align}
Here $\Gamma\approx 4/3$ is the adiabatic index, and $\overline{U}_\mathrm{f}\approx \sqrt{(3/4)\, \kappa_\mathrm{f}\, \alpha}$ is the root-mean-squared fluid velocity. 
The peak frequency, $f_\mathrm{sw}$, is given by
\begin{align}\label{eq:f_sw}
f_\mathrm{sw}= \bigl( 8.9\,\mu\mathrm{Hz} \bigr) \,\frac{1}{v_w}\left(\frac{\beta}{H}\right) \left(\frac{z_\mathrm{p}}{10}\right) \left(\frac{\Tc}{100\GeV}\right) \left(\frac{g_\ast}{100}\right)^{1/6}
	\com
\end{align}
where $z_\mathrm{p} \simeq 10$ is a simulation-derived factor and $g_\ast$ is the effective number of relativistic species.  Using \erefs{eq:rho_d_mat}{eq:T_d} we can write $g_\ast = g_{\ast, \gamma} + g_{\ast, d} (T_d / T_\gamma)^4$.  
The efficiency coefficient $\kappa_\mathrm{f}$ is in general a function of $v_\mathrm{w}$ and $\alpha$, and a numerical fit of $\kappa_\mathrm{f}(v_\mathrm{w}, \alpha)$ is done in \rref{Espinosa:2010hh} for four different scenarios of wall velocity.  
In our calculation we use
\begin{align}\label{eq:kappa_f}
\kappa_\mathrm{f} = \dfrac{\alpha^{2/5}}{0.017+(0.997+\alpha)^{2/5}} \,,
\end{align}
which corresponds to a subsonic wall velocity.

\begin{figure}[h!]
\begin{center}
\includegraphics[width=0.65\textwidth]{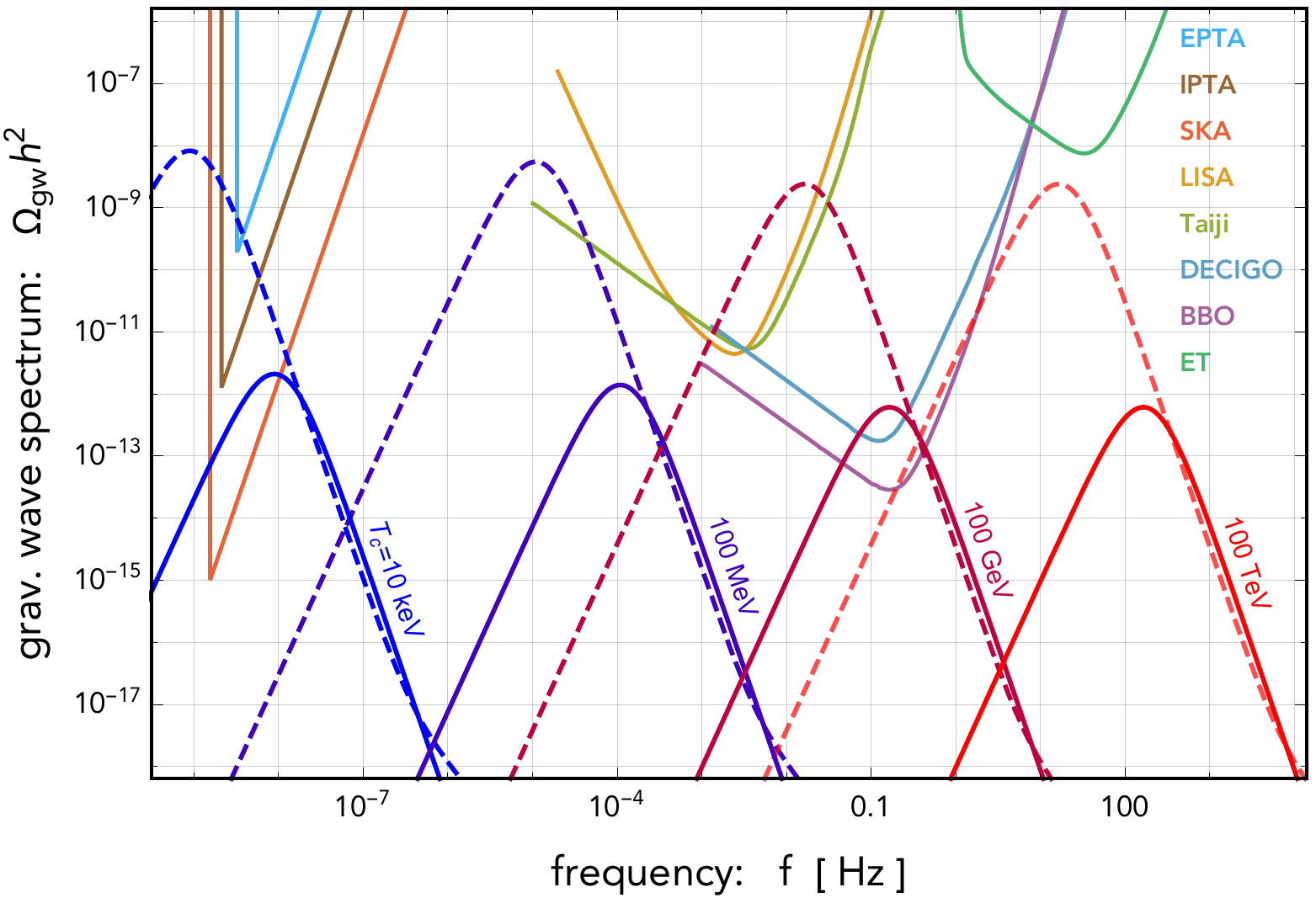}
\caption{\label{fig:GW_spectrum}
We show the GW spectrum that is predicted to arise from a first-order confining phase transition in dQCD along with the projected sensitivities of various future GW interferometer and pulsar timing array experiments~\cite{LISA:sensitivity, Yagi:2011wg, ET:sensitivity, doi:10.1093/nsr/nwx116, Guo:2018npi,0264-9381-30-22-224009,0264-9381-30-22-224010,SKA,Moore:2014lga}.  
We vary the phase transition temperature from $\Tc = 10 \keV$ to $100 \TeV$, and we show $(\alpha, \, \beta/H) = (0.1, \, 10^4)$ (solid) and $(1, \, 10^3)$ (dashed).   
The interferometer sensitivities are calculated using $\Omega_\mathrm{gw} = (2\pi^2 f^3 / 3H_0^2) S_n$ where $S_n^{1/2}$ is the noise amplitude spectral density; often the power-law integrated sensitivity is shown instead, which can be one or two orders of magnitude stronger.  
}
\end{center}
\end{figure}

%=========
Using the formulas above we have calculated the predicted spectrum of gravitational wave radiation, and we present our results in \fref{fig:GW_spectrum}.  
For comparison we also show the projected sensitivities of various GW interferometer observatories and several pulsar timing array experiments.  
In calculating $\Omega_\mathrm{gw} h^2$ we fix $v_w = c_s = 1 / \sqrt{3}$, we assume $T_{\gamma,c} \equiv T_\gamma(t_c) = T_d(t_c) \equiv T_c$, we vary $T_c$ from $10 \keV$ to $100 \TeV$ (corresponding to the different colors), and we choose two combinations of $\alpha$ and $\beta$:  $(\alpha, \, \beta/H) = (0.1, \, 10^4)$ (solid) and $(1, \, 10^3)$ (dashed).  
We also choose $N_d=N_f=3$, which determines $g_\ast = g_{\ast,\gamma} + g_{\ast,d}$ through \eref{eq:rho_d_rad} to be $g_\ast = 3.8, 13.0, 154.25,$ and $154.25$ for $T_c =10 \keV, 100 \MeV, 100 \GeV$, and $100 \TeV$.  
A robust calculation of $\alpha$ and $\beta$ in dQCD is challenging, since the theory becomes strongly coupled at the phase transition.  
Using a low-energy chiral effective description of the phase transition in \aref{app:EFT}, we find that $(\alpha, \, \beta/H)=(0.1, \, 10^4)$ may be typical values; see \fref{fig:GW_params}.  
We also present the GW spectrum for $(\alpha,\, \beta/H)=(1, \, 10^3)$, which is more favorable for detection, to allow for the possibility that the transition is more strongly first order than the chiral effective theory would suggest.   
If the confinement scale is on the lower end, corresponding to $T_c \sim 10 \keV$, then the GW signal will be probed by pulsar timing array observations like  EPTA~\cite{0264-9381-30-22-224009}, IPTA~\cite{0264-9381-30-22-224010} and SKA~\cite{SKA}.  
Alternatively if $T_c \sim 100 \MeV$ to $100 \GeV$ then the GW signal could be accessible to future space-based gravitational wave interferometer experiments like LISA~\cite{LISA:sensitivity}, Taiji~\cite{doi:10.1093/nsr/nwx116, Guo:2018npi}, DECIGO~\cite{Yagi:2011wg}, BBO~\cite{Yagi:2011wg} and ET~\cite{ET:sensitivity}.

%--------------------------------------------
% Cosmic rays from colliding dark quark nuggets
%--------------------------------------------
\subsection{Cosmic rays from colliding and merging dark quark nuggets}\label{sec:cosmic_ray}

%=========
Let us now turn our attention to astro-particle probes of dark quark nuggets in the universe today.  
If a pair of dark quark nuggets were to collide today, some fraction of the initial energy would be liberated as dark radiation (mostly dark mesons), and a new dQN would be formed from the merger.  
If the dark sector has a direct coupling to the Standard Model, the dark mesons may decay into ultra-high energy SM particles, and the observation of these cosmic rays thereby provides a new channel for the indirect detection of dark quark nuggets.

\subsubsection*{Collisions of dark quark nuggets near the Sun}

\begin{figure}[h!]
\begin{center}
\includegraphics[width=0.65\textwidth]{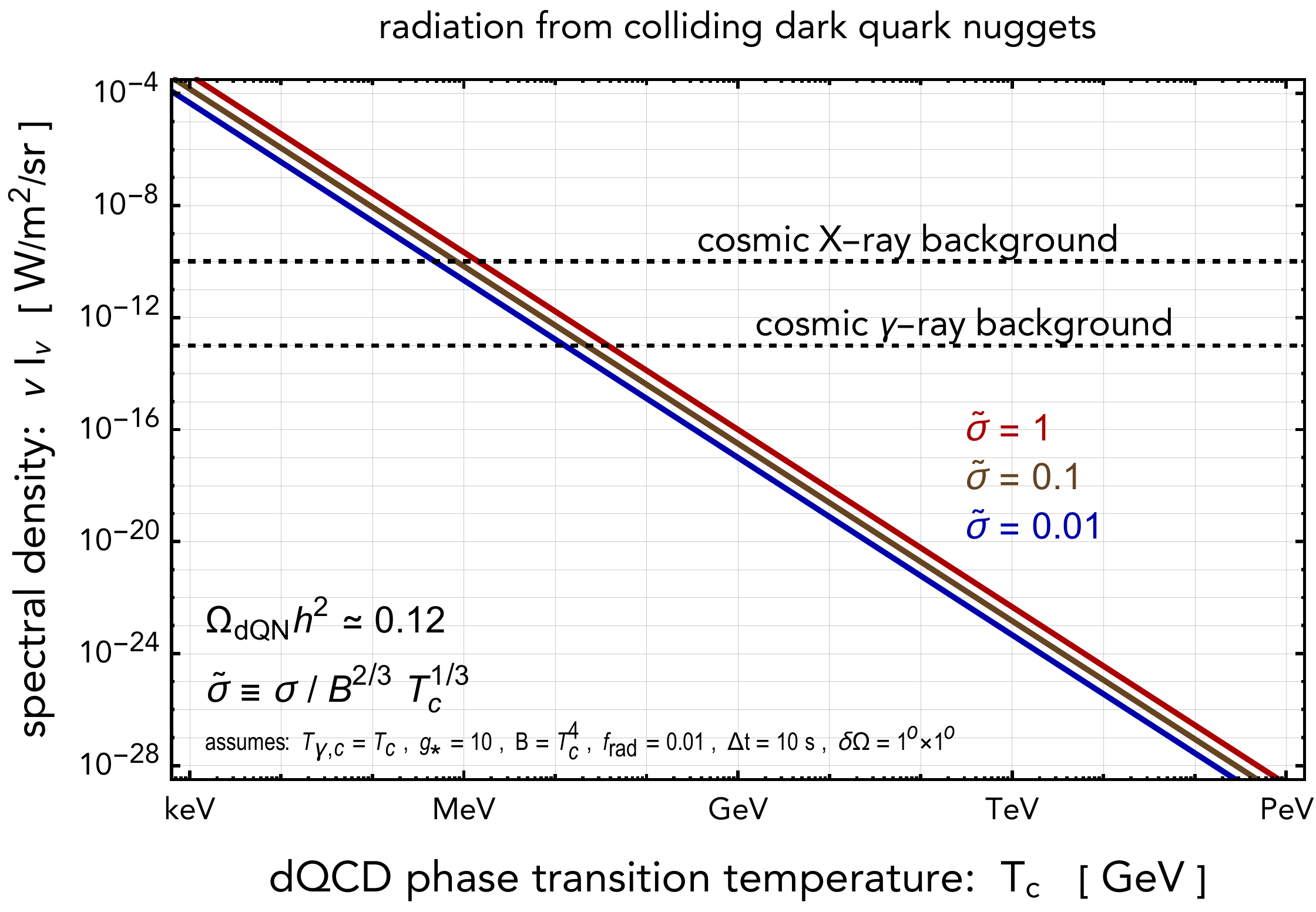} 
\caption{\label{fig:flux_density}
The frequency-weighted spectral density is shown here for colliding dQNs nearby to the Sun.  An angular resolution of $1^\circ \times 1^\circ$ is assumed.  The amplitudes of the cosmic X-ray and $\gamma$-ray backgrounds are shown for comparison; a prediction of $\nu I_\nu$ above this level may be detectable, which corresponds to $T_c \lesssim 10 \MeV$.  
}
\end{center}
\end{figure}

%=========
Let us begin by estimating the rate of dQN collisions nearby to the Sun.  
Here we assume that dark quark nuggets make up all of the dark matter, $\rho_\dQN \approx \rho_\DM \simeq 0.3 \GeV / \mathrm{cm}^3$, and that all nuggets have the same mass and radius: $M_\dQN$ given by \eref{eq:M_dQN_alt} and $R_\dQN$ given by \eref{eq:R_dQN_alt}.  
The rate of dQN collisions per unit volume is estimated as $\gamma_\mathrm{collide} \approx n_\dQN^2 \, v_\dQN \, A_\dQN$ where $n_\dQN = \rho_\DM / M_\dQN$ is the number density of dQNs near the Sun, $v_\dQN = v_\DM \simeq 10^{-3}$ is the typical speed of a dQN in the Milky Way, and $A_\dQN = \pi R_\dQN^2$ is the geometrical cross section of a dark quark nugget.  
(The gravitational enhancement to $A_\dQN$ is negligible.)  
Now consider a spherical region of radius $d$ centered at the Sun.  
The rate of dQN collisions within this region is roughly $\Gamma_\mathrm{collide}(d) \approx \gamma_\mathrm{collide} 4 \pi d^3 / 3$, which evaluates to   
\begin{align}\label{eq:collide-rate}
	\Gamma_\mathrm{collide} 
	& \simeq \bigl( 16 \yr^{-1} \bigr) 
	\left( \frac{B}{(0.1 \GeV)^4} \right)^{-2/3} 
	\left( \frac{\Tc}{0.1 \GeV} \right)^4 
	\left( \frac{\tilde{\sigma}}{0.1} \right)^{-6} 
	\left( \frac{d}{10 \pc} \right)^{3} 
	\per
\end{align}
Similarly we can define a distance $d_\mathrm{yr}$ such that $\Gamma_\mathrm{collide} = 1 \yr^{-1}$, which gives 
\begin{align}
	d_\mathrm{yr} \simeq \bigl( 4.0 \pc \bigr)
	\left( \frac{B}{(0.1 \GeV)^4} \right)^{2/9} 
	\left( \frac{\Tc}{0.1 \GeV} \right)^{-4/3} 
	\left( \frac{\tilde{\sigma}}{0.1} \right)^{2} 
	\per
\end{align}
We estimate the amount of energy liberated during a collision as $2 \times M_\dQN \, v_\dQN^2 / 2$, which is just the kinetic energy of the two incident dQNs.  
Suppose that a fraction $f_\mathrm{rad}$ of this energy goes into visible, SM radiation.  
If the collision takes a time $\Delta t$ to complete, then the corresponding power output is estimated as $P_\mathrm{collide} \approx f_\mathrm{rad} \, M_\dQN \, v_\dQN^2 / \Delta t$, which evaluates to 
\begin{align}
	P_\mathrm{collide} 
	\simeq \bigl( 4.8 \times 10^{-11} \Lsun \bigr)
	\left( \frac{f_\mathrm{rad}}{0.01} \right)
	\left( \frac{\Delta t}{10 \sec} \right)^{-1}
	\left( \frac{\Tc}{0.1 \GeV} \right)^{-3}
	\left( \frac{\tilde{\sigma}}{0.1} \right)^{9/2} 
	\com
\end{align}
where $\Lsun \simeq 3.8 \times 10^{26} \Watt$ is the luminosity of the Sun.  
To assess whether a telescope on Earth could detect this radiation, we assume an angular resolution of $\delta\Omega=1^\circ \times 1^\circ=\left(\pi/180\right)^2\,\mathrm{sr}$.  
Then the frequency-weighted spectral density is estimated as $\nu I_\nu = P_\mathrm{collide}/(d_\mathrm{yr}^2 \,\delta\Omega)$, which evaluates to 
\begin{align}\label{eq:nu_Inu_collide}
	\nu I_\nu \simeq \bigl( 4.1 \times 10^{-15} \frac{\mathrm{W}}{\mathrm{m}^2 \, \mathrm{sr}} \bigr)
	\left( \frac{f_\mathrm{rad}}{0.01} \right)
	\left( \frac{B}{(0.1 \GeV)^4} \right)^{-4/9}
	\left( \frac{\Tc}{0.1 \GeV} \right)^{-1/3}
	\left( \frac{\tilde{\sigma}}{0.1} \right)^{1/2} 
	\per
\end{align}
For comparison, the observed cosmic backgrounds of X-rays and gamma rays run from $\nu I_\nu = 10^{-10} \, \mathrm{W} \,\mathrm{m}^{-2}\,\mathrm{sr}^{-1}$ at $E_\gamma = 10 \keV$ down to $\nu I_\nu = 10^{-13} \, \mathrm{W} \,\mathrm{m}^{-2}\,\mathrm{sr}^{-1}$ at $E_\gamma = 10 \GeV$~\cite{Hill:2018trh}.  
If a dQN collision produces photons with energies in this range, then the signal could be detectable for $B^{1/4} \sim \Tc \lesssim 10 \MeV$.  
This is represented in \fref{fig:flux_density} where we plot $\nu I_\nu$ for different phase transition temperatures. 
The radiation energy is related to the Fermi momentum of the dark quark matter or the phase transition temperature, $T_c$.  
This is similar to a neutron-star merge event, where semi-relativistic neutrons collide with each other to generate energetic photons up to the neutron's kinetic energy.  
For $T_c \gtrsim 10 \keV$, dQN collisions will produce energetic $X$-rays and gamma-rays, which provide transient signals that telescopes can seek out.

%=========

\subsubsection*{Visible radiation from dQN collisions}

%=========
We expect that the collisions of dark quark nuggets will release an enormous number of dark mesons, which may decay into SM-sector particles that could be detected from Earth.  
In this way a dQN collision event may resemble the (less energetic) cousin of a binary neutron star merger.  
The coupling of the dark meson to SM particles depends on unknown UV physics, which we parametrize with the dimension-6 operator, $\overline{\psi}_d \gamma^5 \psi_d \, \overline{\psi} \gamma^5 \psi /\Lambda_\mathrm{UV}^2$, that explicitly breaks the chiral symmetries of the dark quarks, $\psi_d$, and the SM quarks, $\psi$.  
This operator is motivated in \rref{Bai:2013xga} by efforts to relate the dark and visible baryon asymmetries.    
To identify the coupling of the dark mesons, $\pi_d$, we use the relation $\overline{\psi}_d \gamma^5 \psi_d \approx i \, \pi_d\, \langle \overline{\psi}_d \psi_d \rangle / f_{\pi_d} \approx  i \, \pi_d \, \Lambda_d^3 / f_{\pi_d}$, which gives $\Lambda_d^3/(f_{\pi_d} \, \Lambda_\mathrm{UV}^2)\, \pi_d \, i \,  \overline{\psi} \gamma^5 \psi$.  
When the dark meson mass, $m_{\pi_d}$, is far above the SM fermion masses, the two-body decay width is approximated as 
\beqa
\Gamma_{\pi_d} \approx \frac{1}{8\pi}\, \left( \frac{\Lambda_d^3}{f_{\pi_d}\, \Lambda_\mathrm{UV}^2} \right)^2 \, m_{\pi_d} 
\per
\eeqa
We take $\Lambda_d \sim f_{\pi_d}$ to estimate the dark meson lifetime, which is found to be 
\begin{align}
	\tau_{\pi_d} \approx \bigl( 165 \sec \bigr)
	\left( \frac{\Lambda_\mathrm{UV}}{1000 \TeV} \right)^4
	\left( \frac{f_{\pi_d}}{1 \GeV} \right)^{-4}
	\left(\frac{m_{\pi_d}}{0.1 \GeV} \right)^{-1}
	\per
\end{align}
If the dark meson decays into SM particles very quickly, it may allow dQN collisions to provide a visible signal.  
By comparing the mean free path of the dark meson against the typical distance to the source, we find that $c\,\tau_{\pi_d} < d_\mathrm{yr}$ for $\Lambda_d > (0.115 \MeV) (\Lambda_\mathrm{UV} / \mathrm{TeV} )^{36/41}$; here we have taken $m_{\pi_d} = \Lambda_d / 10$, $f_{\pi_d} = \Lambda_d$, $B = \Lambda_d^4$, and $\Tc = \Lambda_d$.  
Similar considerations can yield an estimate of $f_\mathrm{rad}$ in \eref{eq:nu_Inu_collide}.  

\subsubsection*{Mergers of gravitationally-bound dQN systems}

%=========
The preceding calculation only accounts for head-on collisions, but a pair of dark quark nuggets may also form a gravitationally-bound system, which allows them to merge after radiating away excess kinetic energy.  
A pair of dark quark nuggets can form a gravitationally-bound binary system if their relative speed is smaller than their escape speed, $v_\mathrm{rel} < v_\mathrm{esc}$.  
For a pair of nuggets with mass $M_\dQN$, their relative speed at time $t$ is estimated as $v_\mathrm{rel}(t) \approx \sqrt{3 \, T_\gamma(t) / M_\dQN}$, which assumes that the nuggets are in kinetic equilibrium with the SM thermal bath.  
If the nuggets are separated by a distance $D(t)$ at time $t$, then their escape speed at time $t$ is $v_\mathrm{esc}(t) = \sqrt{2 G_N M_\dQN / D(t)}$, where $G_N$ is Newton's constant.  
From \eref{eq:D_init} we recall that the initial nugget separation distance is $D(t_c) = D_\mathrm{init}$, and for non-bounded system this distance grows due to cosmological expansion as $D(t) = D_\mathrm{init} \, \bigl[ a(t) / a(t_c) \bigr] = D_\mathrm{init} \, \bigl[ T_\gamma(t) / \Tc \bigr]^{-1} \bigl[ g_{\ast S}(t) / g_{\ast S}(t_c) \bigr]^{-1/3}$.  
In comparing $v_\mathrm{rel}(t) < v_\mathrm{esc}(t)$ the time-dependence drops out, and we find that a pair of nuggets can be gravitationally-bounded if 
\begin{align}\label{eq:two-body-condition}
	\Tc < T_c^\mathrm{two} 
	\simeq \bigl( 273 \GeV \bigr) 
	\left( \frac{\tilde{\sigma}}{0.1} \right)^{3/2} 
	\per
\end{align}
For larger values of $\Tc$ the nuggets have too much kinetic energy and too little mass to become gravitationally bounded.

%=========
Let us suppose that a pair of nuggets has formed a gravitationally-bound binary system, and we estimate the time $\tau$ that elapses before they merge.    
The orbital radius $r(t)$ decays as the nuggets radiate away energy, according to 
\begin{align}\label{eq:dEdt}
	\frac{dE_\mathrm{grav}}{dt} = \frac{G_N M_\dQN^2}{r^2} \frac{dr}{dt} = P_\mathrm{radiation}(t,r)
	\com
\end{align}
and it reaches zero after a time $\tau$.  
Following \rref{Peters:1964zz} we first estimate the binary system's lifetime that results from gravitational wave emission,\footnote{More generally, the merger time for a pair of masses $m_1$ and $m_2$ is given by $t_\mathrm{GW} = (5/256) \, G_N^{-3} \, D^4 \, (1-e^2)^{7/2} \, [m_1\,m_2\,(m_1 + m_2)]^{-1}$ if the orbital radius and eccentricity are $D$ and $e$, respectively.  To obtain \eref{eq:tau_GW} we take $e = 0$, $D = D_\mathrm{init}$, and $m_1 = m_2 = M_\dQN$.} 
\begin{align}\label{eq:tau_GW}
	\tau_\mathrm{GW} & 
	\simeq \bigl( 3.2 \times 10^{45} \sec \bigr) 
	\left( \frac{\Tc}{0.1 \GeV} \right) 
	\left( \frac{\tilde{\sigma}}{0.1} \right)^{-15/2} 
	\com
\end{align}
which is much larger than even the current age of the universe, $t_0 \simeq 4.32 \times 10^{17} \sec$.

%=========
Next we consider the orbital decay due to the emission of massless dark gluons, which hadronize to form massless dark mesons.    
Since we are not aware of an analytical expression for the double-dark-gluon radiation power, we adapt the corresponding expression for electromagnetic radiation as a rough estimate.  
The power output by a charged particle moving in a circle of radius $r$ is $P_\mathrm{em} = 2\alpha\, \gamma^4/(3r^2)$ where $\gamma$ is the boost factor.  
This motivates us to estimate the two-dark-gluon-radiation power as $P_\mathrm{2dg} \sim \alpha_d^2 / r^2$ for nonrelativistic motion.  
Using this expression in \eref{eq:dEdt} gives $dr/dt \sim \alpha_d^2 / (G_N M_\dQN^2)$, and we estimate the merger timescale as $\tau_\mathrm{merge} \sim D_\mathrm{init} G_N M_\dQN^2 / \alpha_d^2$, which gives 
\begin{align}\label{eq:merge-rate}
	\tau_\mathrm{merge} 
	\simeq \bigl( 2.3 \times 10^{23} \sec \bigr) 
	\left( \frac{\tilde{\sigma}}{0.1} \right)^{21/2} 
	\left( \frac{\Tc}{0.1 \GeV} \right)^{-8} 
	\left( \frac{\alpha_d}{1} \right)^{-2} 
	\per
\end{align}
This merger timescale is longer than the age of the universe today for 
\begin{align}
	\Tc < T_c^\mathrm{merge}
	\simeq \bigl( 0.52 \GeV \bigr) 
	\left( \frac{\tilde{\sigma}}{0.1} \right)^{21/16} 
	\left( \frac{\alpha_d}{1} \right)^{-1/4} 
	\per
\end{align}

%=========
Thus we have developed the following understanding of dQN mergers.  
For models with a high confinement scale, $T_c^\mathrm{two} < \Tc$, the dQNs do not form gravitationally-bound systems, because they have too much kinetic energy and too little mass; consequently, they do not merge.  
For the low confinement scale, $\Tc < T_c^\mathrm{merge}$, the nuggets do form gravitationally-bound systems, but their masses are too large to efficiently radiate away gravitational energy by dark gluon emission and too low to radiate energy by GW emission; again, they do not merge.  
However for the intermediate confinement scale, $T_c^\mathrm{merge} < \Tc < T_c^\mathrm{two}$, the nuggets form gravitationally bound systems soon after they are produced, and our estimates suggest that they merge on a timescale that is short compared to the age of the universe today.  
In this intermediate case the distribution of nugget masses and sizes may be different from the estimates in \sref{sec:nuggets} due to successive mergers.  
One can study the evolution of the mass distribution, and calculate the mass distribution in the universe today, by solving the coagulation equations~\cite{1916ZPhy...17..557S}.  
For instance, if the merger time were mass-independent and much shorter than the age of the universe~\cite{Hayashi:1975xx}, then the solution is a flat mass distribution up to $t_0/\tau_\mathrm{merge} \times M_\dQN$.  
A more precise determination of the mass spectrum after mergers require numerical simulations and will not be explored here.

%--------------------------------------------
% Directly detecting dark quark nuggets at Earth
%--------------------------------------------
\subsection{Directly detecting dark quark nuggets at Earth}\label{sec:direct}

%=========
In this section we briefly discuss the possibility of detecting dQN dark matter in terrestrial experiments on Earth.  
If dark quark nuggets make up all of the dark matter, then their flux at a detector on Earth is given by $\Fcal_\dQN = n_\dQN v_\dQN$ where $n_\dQN = \rho_\DM / M_\dQN$ with $\rho_\DM \simeq 0.3 \GeV / \mathrm{cm}^3$ and $v_\dQN = v_\DM \simeq 10^{-3}$.  
If the scale of the detector is $L$ and it operates for a time $\Delta t$, then the expected number of dQN to pass through the detector is estimated as 
\begin{align}
	\Fcal_\dQN L^2 \Delta t 
	\simeq \bigl( 2.5 \times 10^{-15} \bigr) 
	\left( \frac{\Tc}{0.1 \GeV} \right)^{3} 
	\left( \frac{\tilde{\sigma}}{0.1} \right)^{-9/2} 
	\left( \frac{L}{10 \, \mathrm{m}} \right)^2 
	\left( \frac{\Delta t}{1 \yr} \right) 
	\per
\end{align}
Imposing $1 < \Fcal_\dQN L^2 \Delta t$ leads to a lower bound on the confinement temperature, 
\begin{align}
	\Tc > \bigl( 7.4 \TeV \bigr) 
	\left( \frac{\tilde{\sigma}}{0.1} \right)^{3/2} 
	\left( \frac{L}{10 \, \mathrm{m}} \right)^{-2/3} 
	\left( \frac{\Delta t}{1 \yr} \right)^{-1/3} 
	\per
\end{align}
From \eref{eq:M_dQN_alt} we recall that $\Tc > 10 \TeV$ implies $M_\dQN < 1 \times 10^{20} \GeV \simeq 2 \times 10^{-4} \gram$.  
For $10 \TeV \lesssim \Tc$ the flux of dQNs through a terrestrial detector can be large, which opens up the possibility of discovering dQN dark matter with future observations (see also \rref{Grabowska:2018lnd}).  
Of course, the detection of dQNs requires a direct coupling between the dark and visible sectors, which introduces additional model dependence.

%==================================
% Conclusion
%==================================
\section{Conclusions}\label{sec:conclusion}

%=========
Whereas many studies of macroscopic dark matter are phenomenological in nature, in this article we have endeavored to provide a compelling theoretical framework in which a macroscopic dark matter candidate arises naturally and its properties and interactions may be calculated from first principles.  We have argued that the formation of dark quark nuggets is expected in confining gauge theories that generically admit a first order phase transition and a dark baryon asymmetry.

%=========
Depending on the confinement scale and the magnitude of the dark baryon asymmetry, a nugget's mass and radius may span several orders of magnitude, $M_\dQN \sim 10^{-7} - 10^{23} \gram$ and $R_\dQN \sim 10^{-15} - 10^{8} \cm$, and their cosmological abundance can match that of the dark matter.  Thus dQN dark matter populates a wide swath of the macroscopic dark matter parameter space.  

%=========
\begin{figure}[th!]
\begin{center}
\includegraphics[width=0.60\textwidth]{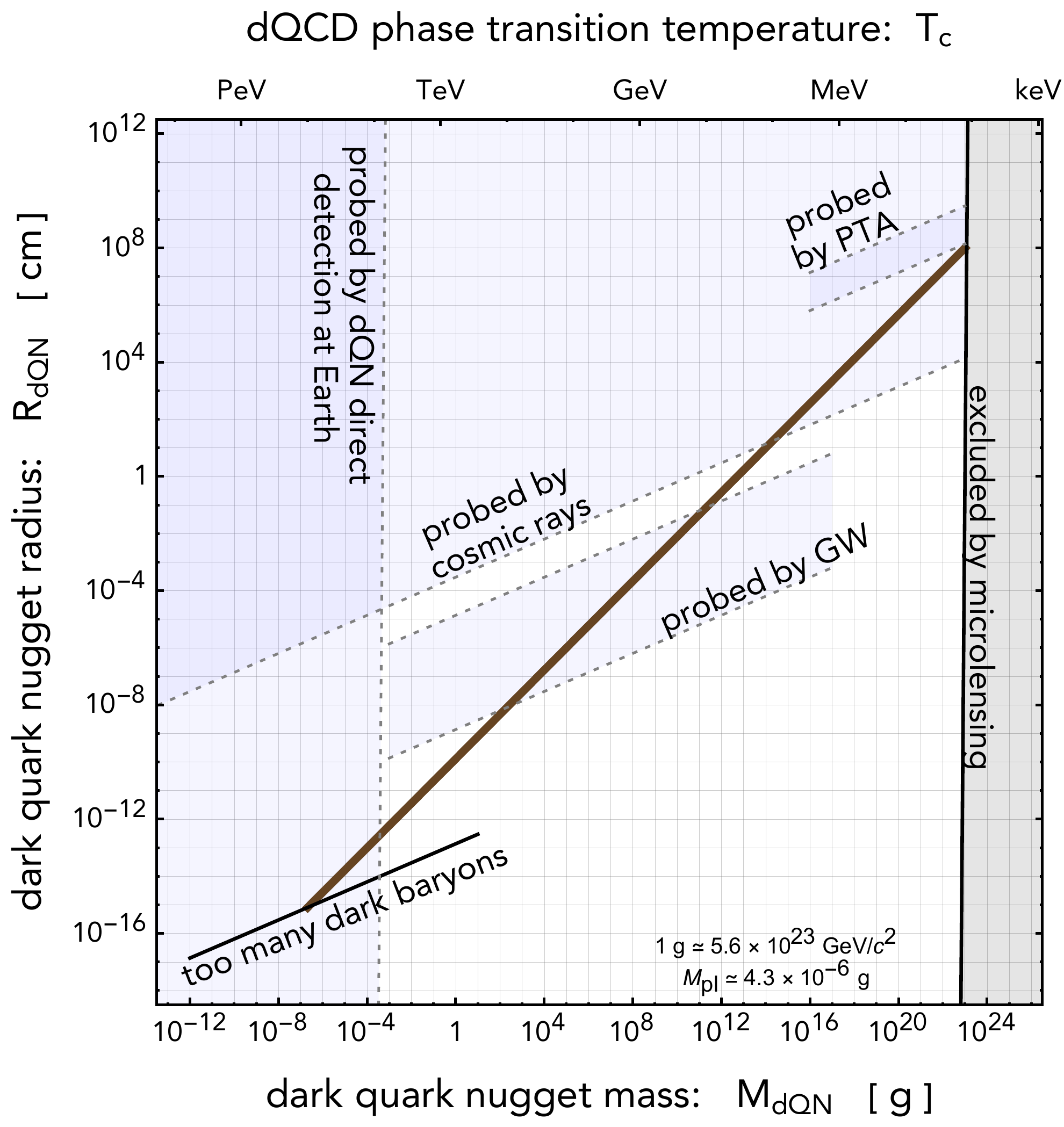} 
\caption{\label{fig:summary}
The predictions and signatures of dark quark nugget dark matter are summarized here.  The predicted dQN mass and radius fall within a couple decades of the brown line (depending on specific choices for model parameters; see also \fref{fig:MdQN}).  Very high-mass nuggets are excluded by searches for microlensing, and we do not expect very low-mass nuggets, because the dQCD model in which they arise also predicts a population of free dark baryons in excess of the dark matter relic abundance.  The first order phase transition, which gives rise to the dQNs, creates a stochastic background of gravitational waves that can be probed by GW interferometry and pulsar timing array observations.  If the theory also admits a direct coupling between the dark and visible sectors, then low-mass nuggets can be probed in laboratories on Earth, while the collisions of high-mass nuggets in the Milky Way halo could be probed through cosmic ray observations.  
}
\end{center}
\end{figure}

%=========
Depending on their mass scale, dark quark nuggets are accessible to a variety of probes, which include gravitational wave radiation, gravitational lensing, cosmic rays, and direct detection on Earth.  
We summarize the probes of dQN dark matter in \fref{fig:summary}.  
In addition the model of $\SU{N_d}$ dark QCD, which gives rise to the dQN studied here, also predicts additional signatures that provide an indirect handle on the physics of dark quark nuggets.  
The formation of dark quark nuggets requires the theory to contain $N_f \geq 3$ flavors of light dark quarks, which become light (and possibly massless) dark mesons after confinement and chiral symmetry breaking.  
If the mass scale of these mesons is below $\sim 1 \eV$ then their presence in the universe is strongly constrained by CMB probes of dark radiation.  
For instance, if $N_d = N_f = 3$ then the predicted dark radiation is at the level of $\Delta N_\mathrm{eff} \simeq 0.21$, which runs into CMB constraints that impose $\Delta N_\mathrm{eff} \lesssim 0.2$ at $95\%$ confidence level, and which can be tested definitively with next-generation CMB-S4 instruments.  
Whereas the dark radiation constraints only rely upon the dark mesons' gravitational influence, a direct coupling between the dark and visible sectors opens the possibility to find evidence for free dark baryons and antibaryons at direct detection experiments on Earth.  
Assuming a vector-vector interaction between dark quarks and SM quarks, we estimate the interaction cross section in \eref{eq:sigma_SI}, and we find $\sigma^\mathrm{SI}_{B_d-n} \sim 10^{-44} \cm^{2}$ if the scale of new physics is $10 \TeV$.  
The sensitivities of current dark matter direct detection experiments like XENON1T are more than adequate to probe these interactions, even if the dark baryons are only a subdominant population of the dark matter.  
Thus the detections of dark radiation and free dark baryons may provide the first clues for the physics of dark QCD and dark quark nugget dark matter.

%=========
Regarding directions for future work, there are several places at which our analysis could be extended and our calculations could be refined.  (1)  We have taken the dark baryon asymmetry to be a free parameter, which may differ from the baryon asymmetry in the visible sector, and it would be useful to investigate how these asymmetries are generated initially in the early universe.  (2)  While the dark and visible sectors may be thermalized in the early universe, this scenario is becoming tightly constrained by CMB limits on dark radiation.  We also consider a scenario in which the two sectors are thermally decoupled, and it would be interesting to study how the two sectors are populated and what interactions control their relative temperatures, which we have taken as a free parameter.  (3)  We have argued that dQN mergers may be frequent for an intermediate mass range, and it would be very interesting to study the effect of these mergers on the dQN mass distribution and the associated observables.  (4)  Our analysis of the observational prospects for colliding dQNs in the Milky Way halo compares the predicted luminosity against the observed diffuse background, but one would like to explore how these transient signals could appear in a specific detector. (5) Finally, the QCD-like gauge theory studied in this paper provides just one example in which macroscopic dark matter can arise from a first-order phase transition in the early universe.  It is worthwhile to explore similar early-universe relics that could be produced in other (supersymmetric) gauge theories or even non-gauge theories.  Overall, we trust that the theory and phenomenology of dark quark nuggets will provide a rich research program in the era of macroscopic dark matter.

%----------------------------------------------------------------
% Acknowledgements
%----------------------------------------------------------------
\subsubsection*{Acknowledgements}
We would like to thank Thomas Appelquist, Jonathan Feng, Patrick Fox, David Weir and Thomas DeGrand for discussions.  
The work of Y.B. and S.L. is supported by the U. S. Department of Energy under the contract DE-SC0017647.  
A.J.L. is supported at the University of Chicago by the Kavli Institute for Cosmological Physics through grant NSF PHY-1125897 and an endowment from the Kavli Foundation and its founder Fred Kavli.  
A.J.L. is supported at the University of Michigan by the US Department of Energy under grant DE-SC0007859. 
This work was performed at the Aspen Center for Physics, which is supported by National Science Foundation grant PHY-1066293. YB also thanks the hospitality of the particle theory group of the University of Chicago and  the Center for Future High Energy Physics at the Institute of High Energy Physics of the Chinese Academy of Sciences.

%----------------------------------------------------------------
% Appendix
%----------------------------------------------------------------
\begin{appendix}

%-------------------------------------------
\section{Low-energy description of the phase transition}\label{app:EFT}

%=========
We can study the phase transition from the low-energy perspective by using a chiral effective theory.  
To describe the phase of broken chiral symmetry, the appropriate dynamical variable is the quark condensate, $\Sigma_{ij} \sim \langle \bar{\psi}_i (1+\gamma_5) \psi_j \rangle$, which transforms as a bi-fundamental under the flavor symmetry group, $\U{N_f}_L \times \U{N_f}_R$.  
We also now specify to $N_f = 3$ flavors for which $\mathrm{det}\,\Sigma$ is cubic in the field and a renormalizable operator.  
The effective theory can be written as~\cite{Bai:2017zhj} 
\begin{align}\label{eq:L_eff}
	\Lscr_\mathrm{eff} 
	= g^{\mu\nu} \, \mathrm{Tr}\bigl( \partial_\mu \Sigma \, \partial_\nu \Sigma^\dagger \bigr) - \Bigl\{ & B - m_\Sigma^2 \, \mathrm{Tr}\bigl( \Sigma \Sigma^\dagger \bigr) - \bigl( \mu_\Sigma \, \det \Sigma + \mu_\Sigma^\ast \, \det \Sigma^\dagger \bigr) 
	\nn & \
	+ \frac{\lambda}{2} \bigl[ \mathrm{Tr}\bigl( \Sigma \Sigma^\dagger \bigr) \bigr]^2 + \frac{\kappa}{2} \, \mathrm{Tr}\bigl( \Sigma \Sigma^\dagger \Sigma \Sigma^\dagger \bigr) \Bigr\}  ~,
\end{align}
where $g^{\mu\nu}$ is the inverse of the metric.  
The five model parameters are the vacuum energy density $B$, the squared mass parameter $m_\Sigma^2$, the dimensionless couplings $\lambda$ and $\kappa$, and the complex mass parameter $\mu_\Sigma$.  
Without loss of generality, it is possible to perform a field redefinition (global phase rotation) that makes $\mu_\Sigma$ real and nonnegative.  

%=========
The symmetry structure of this theory is discussed at length in \rrefs{Pisarski:1983ms,Bai:2017zhj}. In the vacuum where $\langle \Sigma_{ij} \rangle = 0$, the symmetry group is $\SU{N_f}_L \times \SU{N_f}_R\times U(1)_V$.  
In the vacuum where $\langle \Sigma_{ij} \rangle = (f_\Sigma/\sqrt{6}) \, \delta_{ij}$, the symmetry is spontaneously broken to $\SU{N_f}_V \times \U{1}_V$, and the spectrum contains $N_f^2 - 1$ massless Goldstone bosons corresponding to the broken symmetry generators of $\SU{N_f}_A$.  

%=========
To study the phase transition between the symmetric and broken phases, it is convenient to write $\Sigma_{ij} = (\varphi/\sqrt{6}) \, \delta_{ij}$.  
Thus the effective Lagrangian reduces to 
\begin{align}
	\Lscr_\mathrm{eff} 
	= \frac{1}{2} \bigl( \partial_\mu \varphi \bigr)^2 - \Bigl\{ & B - \frac{1}{2} m_\Sigma^2 \varphi^2 - \frac{\mu_\Sigma}{3\sqrt{6}} \varphi^3 + \frac{1}{4} \left( \frac{\lambda}{2} + \frac{\kappa}{6} \right) \varphi^4 \Bigr\} 
	\per
\end{align}
For models with $m_\Sigma^2 > 0$, $\mu_\Sigma > 0$, $\lambda > 0$, and $\kappa > 0$, the scalar potential has its global minimum at $\varphi = f_\Sigma$ where the vacuum expectation value is given by 
\begin{align}
	f_\Sigma = \sqrt{ \frac{m_\Sigma^2}{\lambda/2 + \kappa/6} } \left( \gamma + \sqrt{\gamma^2 + 1} \right) 
	\com
\end{align}
and the dimensionless parameter $\gamma > 0$ is defined by $\gamma \equiv \mu_\Sigma / \sqrt{24 (\lambda/2 + \kappa/6) m_\Sigma^2}$.  
We choose 
\begin{align}
	B = \frac{\bigl( \gamma + \sqrt{\gamma^2 + 1} \bigr)^2 + 2}{12} \, m_\Sigma^2 f_\Sigma^2 
	\com
\end{align}
such that the potential vanishes at $\varphi = f_\Sigma$, and therefore $B$ corresponds to the differential vacuum energy (or pressure) between the phases at $\varphi = 0$ and $\varphi = f_\Sigma$.  

%=========
To study the chiral symmetry breaking phase transition in this model, we calculate the thermal effective potential, $V_\mathrm{eff}(\varphi, T)$, which is the Helmholtz free energy or equivalently the negative pressure of the system.  
In total $\Sigma(x)$ represents $2N_f^2 = 18$ degrees of freedom, which is made transparent by the following parametrization: 
\begin{align}
	\Sigma_{ij} = \frac{\varphi + i \, \phi_I}{\sqrt{6}} \, \delta_{ij} + \Theta^a_R \, (T^a)_{ij} + i \, \Theta^a_I \, (T^a)_{ij}
	\per
\end{align}
The matrices denoted by $T^a$ are the $N_f^2 -1 = 8$ generators of $\SU{N_f} = \SU{3}$.   
The fields $\phi_I(x)$, $\Theta_R^a(x)$, and $\Theta_I^a(x)$ couple to the field $\varphi(x)$ and contribute to the effective potential.  
The one-loop thermal effective potential can be calculated using standard techniques~\cite{Quiros:1999jp}, and by doing so we find 
\begin{align}\label{eq:V_eff}
	V_\mathrm{eff}(\varphi, T) & = B -\frac{1}{2} m_\Sigma^2 \varphi^2 - \frac{\mu_\Sigma}{3\sqrt{6}} \varphi^3 + \frac{1}{4} \left( \frac{\lambda}{2} + \frac{\kappa}{6} \right) \varphi^4 + \! \! \! \sum_{i=\varphi, \phi_I, \Theta_R, \Theta_I} \nu_i \, \frac{T^4}{2\pi^2}\,J_B \bigl[ m^2_i(\varphi, T)/T^2 \bigr] 
	\per
\end{align} 
We have neglected the (zero-temperature, one-loop) Coleman-Weinberg correction~\cite{Coleman:1973jx}, which primarily serves to renormalize the tree-level couplings.  
The thermal correction is expressed as a sum over species that couple to $\varphi$; the multiplicities are $\nu_{\varphi} =  \nu_{\phi_I} = 1$ and $\nu_{\Theta_R} = \nu_{\Theta_I} = 8$; the background-dependent masses are 
\begin{align}\label{eq:effective_masses}
	m^2_{\varphi} & = \left(\frac{5\lambda}{6} + \frac{\kappa}{2}\right)T^2 - m^2_\Sigma - \frac{2}{\sqrt{6}} \mu_\Sigma \, \varphi + 3 \left( \frac{\lambda}{2} + \frac{\kappa}{6} \right) \varphi^2 \, , \nn
	m^2_{\phi_I} & = \left(\frac{5\lambda}{6} + \frac{\kappa}{2}\right)T^2 - m^2_\Sigma + \frac{2}{\sqrt{6}} \mu_\Sigma \, \varphi + \left( \frac{\lambda}{2} + \frac{\kappa}{6} \right) \varphi^2 \,, \nn
	m^2_{\Theta_R} & = \left(\frac{5\lambda}{6} + \frac{\kappa}{2}\right)T^2 - m^2_\Sigma + \frac{1}{\sqrt{6}} \mu_\Sigma \, \varphi + \left( \frac{\lambda}{2} + \frac{\kappa}{2} \right) \varphi^2 \,, \nn
	m^2_{\Theta_I} & = \left(\frac{5\lambda}{6} + \frac{\kappa}{2}\right)T^2 - m^2_\Sigma - \frac{1}{\sqrt{6}} \mu_\Sigma \, \varphi + \left( \frac{\lambda}{2} + \frac{\kappa}{6} \right) \varphi^2
	\, ; 
\end{align}
and the bosonic thermal function is defined by the integral $J_B(y) = \int_0^\infty \ud x \, x^2 \log(1-e^{-\sqrt{x^2+y}})$. 
In the dark QCD model under consideration here, we only keep the contribution to $\varphi$ from light degrees of freedom and ignore the heavy field (e.g., dark baryons) contributions, which are Boltzmann suppressed.

%=========
Around the temperature of the chiral phase transition, the thermal effective potential admits a pair of local minima at $\varphi = 0$ and $\varphi = v_\varphi(T)$, which correspond to the phases of unbroken and broken chiral symmetry, respectively.  
The degeneracy condition, 
\begin{align}\label{eq:Tc_def}
	V_\mathrm{eff}(0,T_c) = V_\mathrm{eff}\left[v_\varphi(T_c), T_c\right]
	\qquad \text{(critical temperature)}
	\com
\end{align}
defines the critical temperature $T_c$ at which the two phases have equal pressure.  
For $T > T_c$ the system is completely in the chiral-unbroken phase, and for $T < T_c$ there is a nonzero probability to nucleate bubbles of the chiral-broken phase.  
Let $S_3(T)$ denote the energy of the static, $\SO{3}$-symmetric critical bubble solution (bounce solution), which can be calculated from $V_\mathrm{eff}(\varphi,T)$ using standard techniques~\cite{Linde:1980tt}, and we provide an analytical approximation below.  
The bubble nucleation rate per unit volume, $\gamma(T) = \Gamma / V$, is written as~\cite{Linde:1980tt}
\begin{align}\label{eq:gamma}
	\gamma \approx \omega \, T^4 \, \left( \frac{S_3}{2\pi T}\right)^{3/2} \, e^{-S_3/T} \,,
\end{align}
where $\omega$ is an order-one, temperature-independent number.  
Nucleated bubbles expand due to the differential vacuum pressure across the phase boundary, but their growth is retarded due to ``friction'' from the plasma~\cite{Fuller:1987ue,Espinosa:2010hh}.  
We assume that the wall quickly reaches a non-relativistic terminal velocity $v_w$, and that the wall is preceded by a shock front that moves at the speed of sound, $v_\mathrm{sh} \approx c_s \simeq 1/\sqrt{3}$~\cite{Kajantie:1992uk}.  
In order to estimate how much time elapses until the shock fronts begin to collide, we let $h(t)$ be the fraction of space that remains in the (unstable) chiral-unbroken phase and outside of a shock front at time $t$.  
This fraction is given by~\cite{Enqvist:1991xw} 
\begin{align}
	h(t) = \mathrm{exp}\Bigl[ - \frac{4\pi}{3} \int^t_{t_c} \! \ud t' \, v_\mathrm{sh}^3\,(t-t')^3\, \gamma(t') \Bigr] \,,
\end{align}
where $t_c$ is the time at which the plasma temperature equals $T_c$.  
We define the fiducial bubble nucleation time $t_n$ by the condition $h(t_n) = 1/e$.  
The integrand is dominated by $t^\prime = t_n$, and we can use the saddle-point approximation to evaluate the integral.  
We first write $\gamma(t^\prime) = \mathrm{exp}{[\ln \gamma(t^\prime)]}$ and then approximate $\ln{\gamma(t^\prime)} \approx \ln{ \gamma(t_n)} +\, (t^\prime - t_n)\, \xi$ where 
\begin{align}\label{eq:xi_def}
	\xi 
	\equiv \frac{d}{dt} \ln \gamma \bigr|_{t_n} 
	= \beta - \frac{3}{2}\, \frac{\beta}{(S_3 / T) \bigr|_{t_n}} + 4 \, \frac{\dot{T}}{T} \Bigr|_{t_n} \,,
\end{align}
and where 
\begin{align}\label{eq:beta_def}
	\beta \equiv - \frac{d(S_3/T)}{dt} \Bigr|_{t = t_n} = \left( \frac{\dot{T}/T}{-H} \right) \left( T \frac{d(S_3/T)}{dT} \right) H \, \biggr|_{t = t_n}
	\per
\end{align}
If the plasma cools due to adiabatic cosmological expansion then $\dot{T}/T = -H - \dot{g}_{\ast S}/3 g_{\ast S} \approx -H$.  
Moreover, typically $(S_3/T)|_{t_n} \gg 1$ and $\beta \gg H$ such that $\xi \approx \beta$.  
Then $h(t_n) = 1/e$ gives 
\begin{align}\label{eq:tn_def}
	1 \ \approx \ \frac{4\pi}{3} \int_{t_c}^{t_n} \! \ud t' \, v_\mathrm{sh}^3 \, (t_n - t')^3\, \gamma(t_n) \, e^{(t^\prime - t_n) \beta} \ \approx \ 8\pi v_\mathrm{sh}^3 \gamma(t_n) \, \beta^{-4} 
	\com
\end{align}
which determines the fiducial bubble nucleation time $t_n$.  
The parameter $\beta$ also provides a fiducial measure of the phase transition duration, since the bubble nucleation rate $\gamma \sim e^{-S_3/T}$ grows by a factor of $e$ on a time scale set by $\Delta t \sim \beta^{-1}$.  
Let $n_\mathrm{nucleations}$ be the average density of bubble nucleation sites (coarse-grained on a scale that's much bigger than the typical inter-site separation) that occur before the phase transition finishes.  
We can estimate the nucleation density as~\cite{Enqvist:1991xw} 
\begin{align}\label{eq:n_nucleations}
	n_\mathrm{nucleations} 
	= \int_{t_c}^{\infty} \! \ud t^\prime \, \gamma(t^\prime) \, h(t^\prime) 
	\, \approx \, \bigl( 8 \pi v_\mathrm{sh}^3 \beta^{-3} \bigr)^{-1}  \,,
\end{align}
where we have used the saddle point approximation to evaluate the integrals.  
Now all that remains is to calculate the bounce energy, $S_3(T)$, and evaluate $\beta$ with \eref{eq:beta_def}.  

\begin{figure}[t]
\begin{center}
\includegraphics[width=0.65\textwidth]{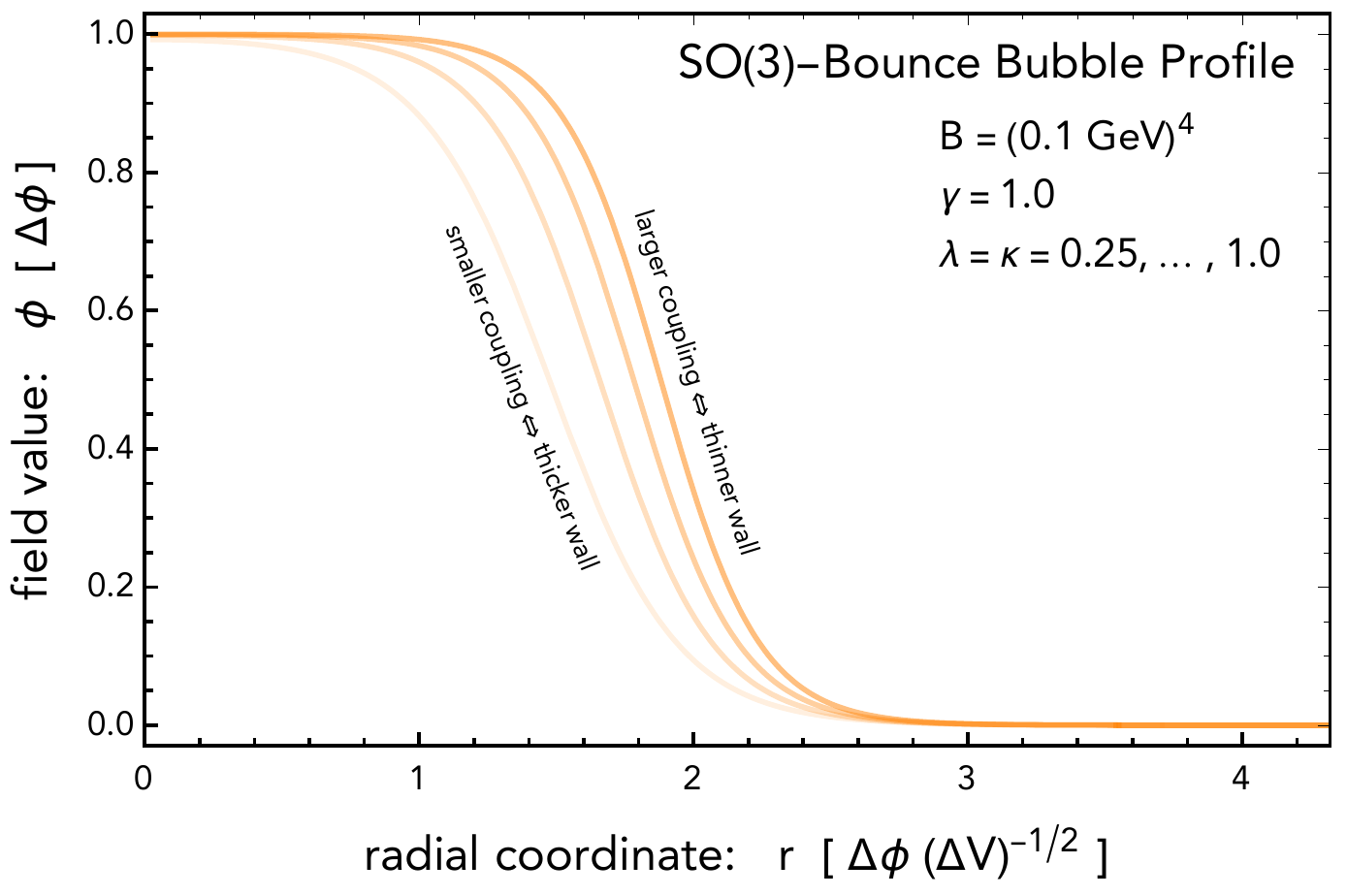} 
\caption{\label{fig:thin_wall}
We show the bubble solution's profile functions for various values of the dimensionless, quartic couplings ($\lambda$ and $\kappa$) in the chiral effective theory of \eref{eq:L_eff}.  As we consider more strongly coupled theories, larger $\lambda$ and $\kappa$, we find that the bubble solution becomes thin walled.  The profile function is scaled by $\Delta \phi = v_\varphi(T_n)$ and $\Delta V = V_\mathrm{eff}(0,T_n) - V_\mathrm{eff}\left[v_\varphi(T_n),T_n\right]$.  
}
\end{center}
\end{figure}

%=========
Using direct numerical evaluation, we have calculated the bounce solution for the thermal effective potential in \eref{eq:V_eff}.  
As we raise the size of the couplings, $\lambda$ and $\kappa$, we find that the bounce solution takes the form of a thin-walled bubble.  
This result is illustrated in \fref{fig:thin_wall}.  
We have studied a slice of parameter space along which $B = (0.1 \GeV)^4$, $\gamma = 1.0$, and $\lambda = \kappa$ varies from $0.25$ to $1.0$.   
Thin-walled bubbles result when $T_n \lesssim T_c$, and this occurs for large couplings because the effective potential responds ``rapidly'' to changes in temperature.  
For instance the thermal mass terms in \eref{eq:effective_masses} imply that $V_\mathrm{eff} \sim (5\lambda/6 + \kappa/2) T^2 \varphi^2$.

%=============

%=========
For a thin-wall bubble the bounce action can be approximated as~\cite{Linde:1980tt,Fuller:1987ue} 
\begin{align}\label{eq:S3ovT_thin_wall}
	\frac{S_3}{T} \approx \frac{16\pi}{3}\frac{T_c\,\sigma^3}{L^2\,(T_c - T)^2} 
	\qquad , \ (T < T_c)
\end{align}
where $L$ is the latent heat of the phase transition and $\sigma$ is the bubble's surface tension at the time of its nucleation.  
Parametrically the latent heat is set by the differential vacuum pressure, $B$, and \rref{Fuller:1987ue} estimates $L \approx 4B$, which we will now adopt as a fiducial reference point.  
Using \eref{eq:S3ovT_thin_wall} we evaluate the bubble nucleation rate, given by \eref{eq:gamma}, and the parameter $\beta$, defined in \eref{eq:beta_def}.  
Then by solving \eref{eq:tn_def} we obtain the fiducial bubble nucleation temperature, $T_n$, and we calculate the dimensionless supercooling parameter, $\eta_n \equiv (T_c - T_n) / T_c$, which is found to be 
\begin{align}\label{eq:eta_approx}
	\eta_n & \approx \sqrt{\frac{\pi}{3}} \, \tilde{\sigma}^{3/2} \left[ \log \left( \frac{9 \sqrt{3}}{4\sqrt{2} \, \pi^3} \ \frac{\omega \, T_c^4 \, v_\mathrm{sh}^3}{H_n^4 \, \tilde{\sigma}^{15/2}} \ \eta_n^9 \right) \right]^{-1/2} 
	\per
\end{align}
Here we have introduced the dimensionless tension parameter, $\tilde{\sigma} \equiv \sigma / (B^{2/3} T_c^{1/3})$, which affects the rate of bubble nucleation through \eref{eq:S3ovT_thin_wall} and controls the amount of supercooling through \eref{eq:eta_approx}.  
The Hubble parameter at the fiducial bubble nucleation time, $H_n = H(t_n)$, depends on the dominant energy component of the universe at this time.  
To be general, we allow that the temperature of the plasma in the (dark) sector undergoing the phase transition may be different from the temperature in the (visible) sector.  
By writing the energy densities of radiation in the dark and visible sectors as $\rho_d(t) = (\pi^2/30) g_{\ast,d}(t) \, T_d(t)^4$ and $\rho_\gamma(t) = (\pi^2/30) g_{\ast,\gamma}(t) \, T_\gamma(t)^4$, the Hubble parameter is given by $3 \Mpl^2 H_n^2 = \rho_d(t_n) + \rho_\gamma(t_n) = (\pi^2/30) g_\ast(t_n) T_\gamma(t_n)^4$ where $g_\ast(t) = g_{\ast,\gamma}(t) + g_{\ast,d}(t) (T_d / T_\gamma)^4$.  
Using this expression, the supercooling factor in \eref{eq:eta_approx} becomes 
\begin{align}\label{eq:eta_n_thin_wall}
	\eta_n & \simeq 0.0028
	\left( \frac{\tilde{\sigma}}{0.1} \right)^{3/2}
	\biggl( 1 + 
	0.027 \log \frac{\tilde{\sigma}}{0.1}
	+ 0.014 \log \frac{T_c}{0.1 \GeV} 
	\nn & \qquad \qquad \qquad \qquad \qquad \quad 
	+ 0.029 \log \frac{T_\gamma(t_n)}{T_c}
	- 0.033 \log \frac{\eta_n}{0.0028} 
	\biggr)
	\per 
\end{align} 
For the numerical estimate we have fixed $v_\mathrm{sh} = 1/\sqrt{3}$, $\omega = 1$, and $g_{\ast}(t_n) = 10$.   
A value of $\eta_n \ll 1$ implies that the phase transition occurs after little supercooling, and $T_n$ is just slightly below $T_c$.  
The parameter $\beta$ is given by \eref{eq:beta_def}, which evaluates to 
\begin{align}
	\beta / H_n 
	\, = \, \frac{2\pi}{3} \frac{1 - \eta_n}{\eta_n^3} \, \tilde{\sigma}^3 
	\, \simeq \, \bigl( 1.0 \times 10^5 \bigr) \left( \frac{\tilde{\sigma}}{0.1} \right)^{-3/2} \left( \frac{\eta_n}{0.0028 \, \bigl( \frac{\tilde{\sigma}}{0.1} \bigr)^{3/2}} \right)^{-3}  
	\per
\end{align}
The density of bubble nucleation sites is given by \eref{eq:n_nucleations}, which evaluates to 
\begin{align}\label{eq:n_nucleations_numerical}
	n_\mathrm{nucleations} \, H_n^{-3} 
	\, = \, \bigl( 2.1 \times 10^{14} \bigr) \left( \frac{\tilde{\sigma}}{0.1} \right)^{-9/2} \left( \frac{\eta_n}{0.0028 \, \bigl( \frac{\tilde{\sigma}}{0.1} \bigr)^{3/2}} \right)^{-9}  
	\per
\end{align}
This corresponds to roughly $10^{14}$ nucleation sites per Hubble volume, $V_H \sim H_n^{-3}$.  
Note that $n_\mathrm{nucleations}$ is very sensitive to the amount of supercooling, $\eta_n$, and to the model parameters through $\tilde{\sigma}$.  
If the dark sector radiation energy density is subdominant to the visible sector radiation, then $n_\mathrm{nucleations}$ is insensitive to the temperature in the dark sector, but instead $n_\mathrm{nucleations} \sim H_n^3 \sim T_\gamma^6 / \Mpl^3$.

%=========
Finally it is useful to define a dimensionless parameter, 
\begin{align}\label{eq:alpha_def}
	\alpha \equiv \frac{V_\mathrm{eff}(\varphi = 0, T=0) - V_\mathrm{eff}\left[v_\varphi(T_n), T=0\right]}{(\pi^2/30) \, g_{\ast}(t_n) \, T_\gamma(t_n)^4} 
	\com
\end{align}
that measures the vacuum energy released during the phase transition and controls the strength of the resulting stochastic gravitational wave background.  
The numerator of \eref{eq:alpha_def} is the difference in the vacuum energies between the symmetric ($\varphi = 0$) and broken phases [$\varphi = v_\varphi(T_n)$] at the fiducial bubble nucleation temperature, $T_n$; its value is bounded from above by $B$, the differential vacuum pressure at zero temperature.    
Using the thermal effective potential described above, we have numerically evaluated $\alpha$ and $\beta$, and the results are shown in \fref{fig:GW_params}.  
In evaluating $\alpha$ we assume that the dark and visible sectors are at the same temperature, $T_\gamma(t_n) = T_d(t_n)$.  

%=============
\begin{figure}[bh!t]\label{fig:GW_params}
\begin{center}
\includegraphics[width=0.47\textwidth]{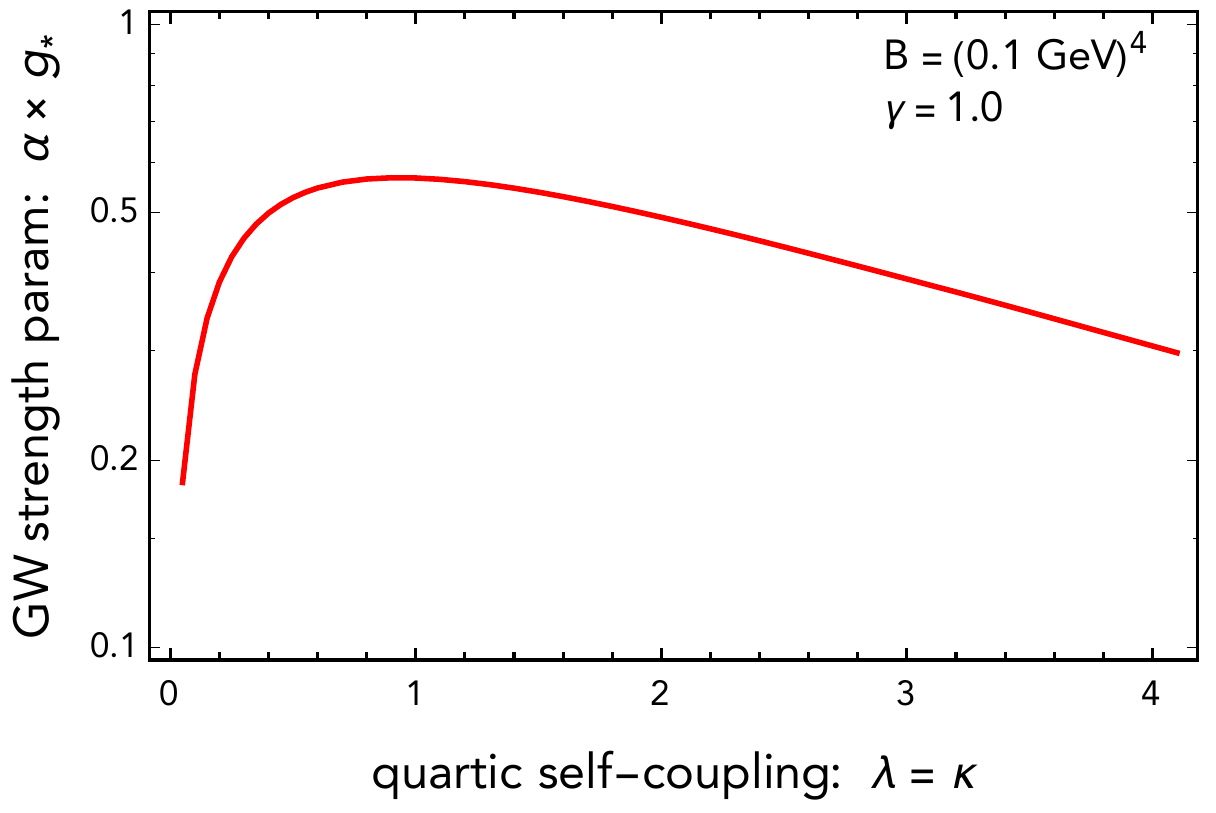} \hfill
\includegraphics[width=0.49\textwidth]{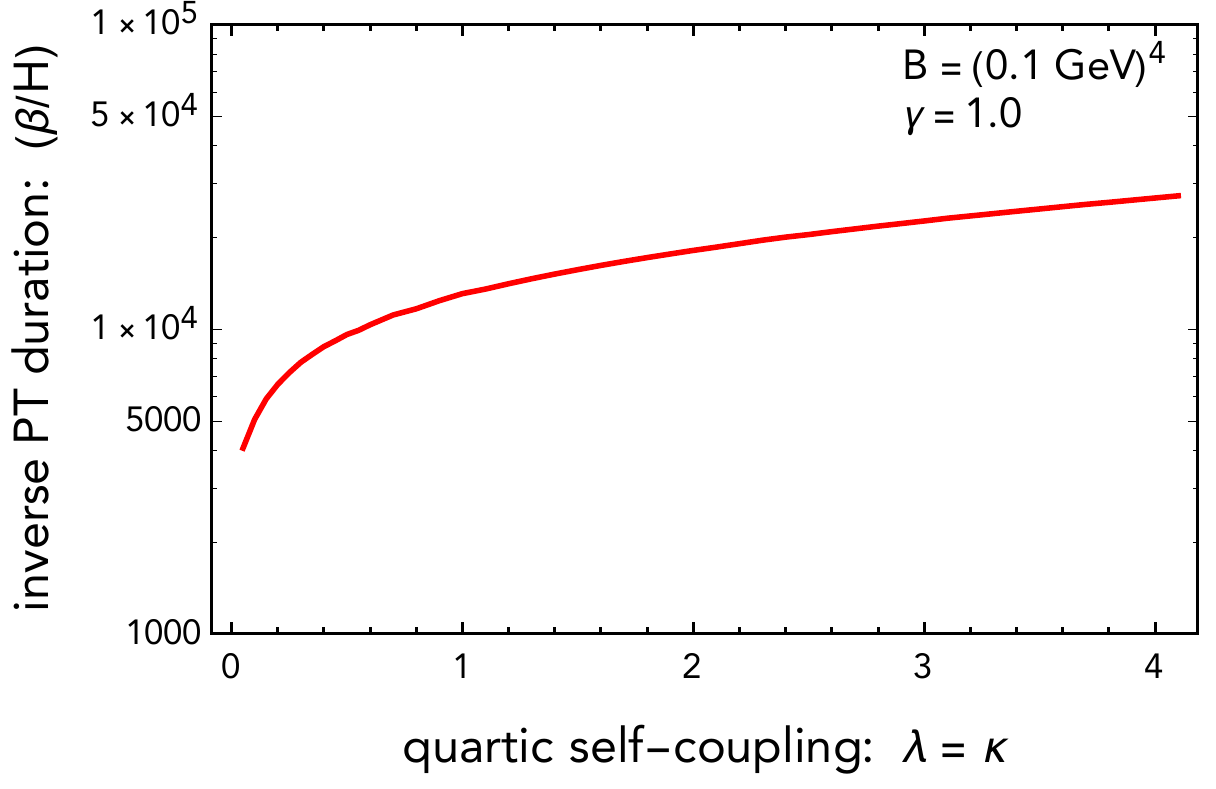} 
\caption{\label{fig:GW_params}
The parameters $\alpha$ and $\beta/H$ that feed into the gravitational wave spectrum are shown here as a function of the dimensionless couplings $\lambda = \kappa$. 
}
\end{center}
\end{figure}

\end{appendix}

%----------------------------------------------------------------
% References
%----------------------------------------------------------------
\bibliographystyle{JHEP}
\bibliography{refs--dark_quark}

\end{document}